\documentclass{aa}
\bibpunct{(}{)}{;}{a}{}{,} 
\usepackage{verbatim}
\usepackage[caption=false]{subfig}
\usepackage{mathrsfs}
\usepackage{bm}
\usepackage{chngpage}
\usepackage[justification=centering]{caption}
\usepackage{txfonts}

\begin{document}

   \title{Evolution Patterns of the Peak Energy in the GRB Prompt Emission}

   \author{Hao-Xuan Gao\inst{1}
          \and
          Jin-Jun Geng\inst{2}
          \and
          Yong-Feng Huang\inst{3,4}
          }

   \institute{School of Physics, Nanjing University, Nanjing 210023, People's Republic of China\\
         \and
             Purple Mountain Observatory, Chinese Academy of Sciences, Nanjing 210023, People's Republic of China\\
             \email{jjgeng@pmo.ac.cn}
         \and
             School of Astronomy and Space Science, Nanjing University, Nanjing 210023, People's Republic of China\\
         \and
             Key Laboratory of Modern Astronomy and Astrophysics (Nanjing University), Ministry of Education, Nanjing 210023, People's Republic of China
             }

   \date{Received Month xx, yyyy; accepted Month xx, yyyy}


   \abstract
   {There are two different evolution patterns of the peak energy ($E_\text{p}$) exhibited during
   the prompt emission phase of gamma-ray bursts (GRBs), i.e., hard-to-soft and intensity-tracking,
   of which the physical origin remains unknown.
   Except for low-energy indices of GRB prompt spectra, the evolution patterns of $E_\text{p}$
   may be another crucial indicator to discriminate radiation mechanisms (e.g., synchrotron or photosphere) for GRBs.}
   {We explore the parameter space to find conditions that could generate different evolution patterns of the peak energy
   in the framework of synchrotron radiation.}
   {We have developed a code to calculate the synchrotron emission from a simplified shell numerically, considering three cooling processes
   (synchrotron, synchrotron self-Compton (SSC), and adiabatic) of electrons, the effect of decaying magnetic field,
   the effect of the bulk acceleration of the emitting shell,
   and the effect of a variable source function that describes electrons accelerated in the emitting region.}
   {After exploring the parameter space of the GRB synchrotron scenario,
   we find that the intensity-tracking pattern of $E_\text{p}$ could be achieved in two situations.
   One is that the cooling process of electrons is dominated by adiabatic cooling or SSC+adiabatic cooling at the same time.
   The other is that the emitting region is under acceleration in addition to the cooling process being dominated by SSC cooling.
   Otherwise, hard-to-soft patterns of $E_\text{p}$ are normally expected.
   Moreover, a chromatic intensity-tracking pattern of $E_\text{p}$ could be induced by the effect of a variable source function.}
   {}


   \maketitle

\section{Introduction}

Gamma-ray bursts (GRBs), the most energetic stellar explosions in the universe,
have been observed and studied for several decades.
The prompt emission of GRBs typically lasts for less than one second to several minutes
and consists of several spikes in the lightcurve,
which is thought to come from the relativistic jet launched from the central compact remnant~\citep{Blandford77,Eichler89,Piran04,Kumar15}.
There are still several fundamental issues for GRBs that remain unsolved, e.g.,
the composition of GRB jets, the radiation mechanism producing GRBs,
and the distance of the emitting region from the central engine~\citep{Zhang18}.

The spectrum of the prompt emission is often described by the so-called Band function empirically~\citep{Band93}.
Except for the Band component, evidence has shown that multiple spectral components
exist in some GRBs, such as an additional power-law component~\citep{Abdo09,Ackermann11} in the high-energy band
or the thermal component~\citep{Ghirlanda03,Ryde05,Ryde09,Ryde10,Pe'er11,Guiriec11,Guiriec13,Lazzati13,Iyyani15,Li19b,Li21b}.
Both synchrotron radiation of electrons~\citep[e.g.,][]{Meszaros94,Tavani96,ZhangB11} and photospheric emission~\citep[e.g.,][]{Rees05,Pe'er06,Ryde11}
have been proposed to explain the GRB prompt emission.
However, it is still under hot debate that which radiation mechanism is responsible for most GRB spectra.

The Band function is characterized by three parameters, the low-energy and
high-energy photon spectral indices ($\alpha$ and $\beta$) as well as the peak energy ($E_\text{p}$).
The ``synchrotron line of death problem''~\citep{Preece98} involving the statistics of $\alpha$
has been proposed to be against the synchrotron origin for GRBs,
which is later eased by detailed treatment on cooling processes of electrons~\citep{Derishev01,Daigne11,Uhm14,ZhaoXH14,Geng18b}.
Meanwhile, some efforts have been made on the revision of the method that observed spectra can be modeled~\citep{ZhengW12,Oganesyan17,Oganesyan18,Ravasio18}.
It is shown that there is a low-energy break in the prompt spectrum of some long GRBs.
The low-energy break is described by two power-law indices $\alpha_{1}$ and $\alpha_{2}$,
which have distributions centered around -2/3 and -3/2, respectively, and agree with the predictions of the synchrotron model.
Notably, this additional spectral break at low energies was also confirmed in some short GRBs through a systematic search~\citep{Ravasio19}.
Moreover, \citet{Burgess20} suggest that synchrotron spectra from electrons in evolving (fast-to-slow) cooling regimes
are capable of fitting $\sim$95\% of all time-resolved spectra of the brightest long GRBs observed by
gamma-ray burst monitor (GBM: 8 keV - 40 MeV) on board the NASA {\it Fermi Gamma-Ray Observatory},
by comparing the theoretical results with observed data directly.

Thus $\alpha$ may not be a good indicator to justify the radiation mechanism~\citep{Burgess15,Iyyani15,Meng18}.
It was found that there are two distinct evolution patterns for $E_\text{p}$, i.e., hard-to-soft~\citep{Norris86}
and intensity-tracking\citep{Golenetskii83,Bhat94,Kargatis94,Ford95,Preece00,Lu10,Ghirlanda11,Lu12,Li19a,Li21b}.
In addition, \citet{Hakkila11} show that all single pulses GRBs could have the hard-to-soft evolution pattern.
Although $E_\text{p}$ is also derived from the Band function,
its evolution pattern with a burst may represent an intrinsic trend rather than an artificial effect like $\alpha$.
Revealing the enigma of these distinct patterns may provide vital clues for understanding the
physical mechanism of the GRB prompt emission.

The evolution patterns of $E_\text{p}$ were studied in different dissipation/radiative processes.
\citet{ZhangB11} proposed that the GRB prompt emission comes from sudden discharge of magnetic energy through turbulent magnetic reconnection,
in which a hard-to-soft evolution of $E_\text{p}$ throughout a pulse is predicted.
In a sophisticated quasi-thermal GRB photosphere scenario, an intensity-tracking pattern for $E_\text{p}$ could be naturally produced
but the hard-to-soft pattern requires contrived physical condition~\citep{Deng14}.
\citet{Beniamini16} discussed the temporal/spectral properties of prompt emission from magnetic reconnection.
They found that magnetic reconnections could account for the hard-to-soft pattern and
the intensity-tracking pattern for different comoving Lorentz factors of the field lines.
However, the evolution of the electron spectra is not incorporated in this work.

In prevailing synchrotron models, the prompt emission comes from a relativistic jet/outflow launched from
the central compact object~\citep[e.g.,][]{Bromberg18,Kathirgamaraju19,Geng19}.
For a baryon-dominated jet, internal shocks could efficiently dissipate the relative kinetic energy of
colliding shells into relativistic electrons~\citep{Narayan92,Rees94,Piran99}.
If the jet is Poynting flux-dominated, turbulent magnetic reconnections are expected to accelerate electrons in the jet~\citep{Drenkhahn02b,Lyutikov03,ZhangB11}.
\citet{Uhm18} considered a simple physical scenario, in which a thin relativistic spherical shell expands
while emitting photons isotropically in the shell.
Assuming that the shell is under bulk accelerating~\citep{Uhm16,Jia16} and that photons emitted obeys
an ad-hoc spectral function without specifying its physical origin,
two evolution patterns of $E_\text{p}$ could then be reproduced.
However, a robust conclusion could be achieved only if
the modeling of the energy spectrum of electrons is spontaneously included in the calculations.
Here, we will focus on the synchrotron scenario and numerically calculate the evolution of the electron distribution,
to test the self-consistency of synchrotron origin for the $E_\text{p}$ evolution.

In this study, we investigate the conditions that make the evolution patterns of $E_\text{p}$ in the framework of synchrotron radiation.
Three main cooling processes (i.e., adiabatic, synchrotron, and synchrotron self-Compton (SSC)) for electrons,
and the effect of a decaying magnetic field are coherently considered in the calculation of electron distributions~\citep{Geng18b,ZhangY19}.
Also, other possible effects like the dynamics of the shell~\citep{Uhm16,Li21a}, and the evolving pre-description of accelerated electrons are included in our investigation.
A brief description of our synchrotron scenario is exhibited in Section 2.
Our numerical results are presented in Section 3,
which include three distinct physical effects (3.1-3.3).
To test the applicability of our results, two short GRBs, i.e., GRB 090227B and GRB 110529A are modeled (3.4).
Moreover, as another possible way to distinguish different radiation mechanisms,
the numerical results of the prompt optical emission are presented (3.5).
Finally, in Section 4, we summarize and discuss our results.
Some relevant formulations are given in the appendix.

\begin{table*}
\caption{Parameters of numerical calculations.}
\label{Table:results}
\centering
\begin{tabular}{c c c c c c c c c}
\hline\hline
Model & $\Gamma$ & $\gamma^{\prime}_{m}$ & $B^{\prime}_{0}$    & $N^{\prime}_{\text {inj}}$  & $R_{0}$  & $s$  & $g$    \\

      &          & $(10^{4})$    &($10^{2}$ G) &$(10^{47} \text {s}^{-1}$) & (cm)\\
\hline
$\text {C} 1 \text {R} 14  \Gamma 240 $   & 240   & 2    & 2     & 4    &  $10^{14}$   \\
$\text {C} 1 \text {R} 15  \Gamma 120 $   & 120   & 10   & 1     & 16   &  $10^{15}$   \\
$\text {C} 2 \text {R} 14  \Gamma 240 $   & 240   & 2    & 2     & 64   &  $10^{14}$   \\
$\text {C} 2 \text {R} 15  \Gamma 460 $   & 460   & 5    & 1     & 2    &  $10^{15}$   \\
$\text {C} 3 \text {R} 14  \Gamma 1900 $  & 1900  & 1.4  & 3     & 1.2  &  $10^{14}$   \\
$\text {C} 3 \text {R} 15  \Gamma 1900 $  & 1900  & 1.4  & 3     & 1.2  &  $10^{15}$   \\
$\text {C} 4 \text {R} 14  \Gamma 460  $  & 460   & 5    & 0.8   & 4    &  $10^{14}$   \\
\hline\hline
$\text {A}  \text {R} 14  \Gamma 200 $   & 200   & 3.5  & 0.6   & 20   &  $10^{14}$   & 0.35\\
$\text {A}  \text {R} 14  \Gamma 150 $   & 150   & 3.5  & 0.6   & 20   &  $10^{14}$   & 0.55   \\
$\text {A}  \text {R} 14  \Gamma 80  $   & 80    & 3.5  & 1.0   & 80   &  $10^{14}$   & 1.0   \\
$\text {A}  \text {R} 15  \Gamma 230 $   & 230   & 5    & 2     & 8    &  $10^{15}$   & 1.0   \\
\hline\hline
$\text {AG} 1 \text {R} 14  \Gamma 300 $   & 300   & 4.5  & 0.6   & 32   &  $10^{14}$   & 0.0   &  0.7       \\
$\text {AG} 2 \text {R} 14  \Gamma 150 $   & 150   & 5.0  & 0.6   & 20   &  $10^{14}$   & 0.55  &  0.5       \\
$\text {AG} 1 \text {R} 15  \Gamma 250 $   & 250   & 5    & 2     & 8    &  $10^{15}$   & 0.0   &  1.0       \\
$\text {AG} 2 \text {R} 15  \Gamma 230 $   & 230   & 5    & 2     & 8    &  $10^{15}$   & 1.0   &  1.0       \\
\hline 
\end{tabular}
\tablefoot{
Three distinct groups of cases are listed. In the fist group, letter C in the model name denotes that the shell is moving at a constant Lorentz factor,
the number (1-4) means the cooling process is dominated by synchrotron cooling, SSC cooling, adiabatic cooling and SSC+adiabatic cooling, respectively.
In the second group, letter A denotes that the shell is moving at an increasing Lorentz factor, and $s$ here describes the index of bulk acceleration law.
In the third group, letter G means that a broken power-law profile of $\gamma_{\mathrm{m}}^{\prime}(R)$ is considered,
while the acceleration effect is also taken into account (letter A).
And the index $g$ describes the evolution of $\gamma_{\mathrm{m}}^{\prime}(R)$, which could refer to Equation (\ref{gm}).}
\label{models}
\end{table*}

\section{Physical scenario}

Let's consider a relativistic jet moving from an initial radius of $R_{0}$ with a bulk Lorentz factor of $\Gamma$ and emitting electrons in it.
We trace the evolution of the distribution of electrons and derive observed synchrotron flux density $F_{\nu_{\mathrm{obs}}}$ from these electrons.
The electrons are injected into the co-moving background magnetic field of $B^{\prime}$
(the superscript prime $\prime$ denotes the quantities in the comoving frame hereafter),
and the evolution of electrons could be obtained by solving the continuity equation in the energy space.
The detailed procedure can be found in Appendix A and the numerical method is the same as that in \citet{Geng18b}.
The distribution of injected electrons is assumed to be a power-law of
$Q (\gamma_{\rm e}^{\prime},t^{\prime}) = Q_0 (t^{\prime}) (\gamma_{\rm e}^{\prime} / \gamma_{\rm m}^{\prime})^{-p}$
for $\gamma_{\rm e}^{\prime} > \gamma_{\rm m}^{\prime}$, where $Q_0$ is related to the
injection rate by $N_{\rm inj}^{\prime} = \int_{\gamma_{\rm m}^{\prime}}^{\gamma_{\rm max}^{\prime}}
Q (\gamma_{\rm e}^{\prime},t^{\prime}) d \gamma_{\rm e}^{\prime}$
\footnote{$\gamma_{\rm max}^{\prime}$ is the maximum Lorentz factor of electrons and is
given by the approximation $\gamma_{\rm max}^{\prime} \simeq 10^8 \left(\frac{B^{\prime}}{1~\mathrm{G}} \right)^{-0.5}$
\citep{Huang00}.}.
The electron spectral index $p$ in shock theories ranges from 2.3 to 2.8~\citep{Kirk00,Sironi09a}.
Quite large values ($p>3$) in some GRBs are obtained through synchrotron fitting to the GRB spectra,
making it a challenge to standard simulations in shocks and magnetic reconnection~\citep{Oganesyan19,Ronchi20,Burgess20}.
However, the high-energy part of the spectrum is rarely constrained because GBM data are missing in most cases.
Also, the evolution of the low-energy electron distribution mainly depends on its cooling process
rather than $p$ when high-energy electrons are continuously injected.
Here, an intermediate value $p = 2.7$ is commonly adopted in view of particle simulations and observations,
and this has little influence on our conclusions.
The temporal structure of a GRB pulse consists of a fast rise and a shallow decay.
To mimic such a GRB pulse, the injection of electrons into the shell is ceased at a turn-off time $t_{\text {off}}$.

The Lorentz factor of the electrons that radiate at the GRB
spectral peak energy is roughly $\gamma_{\rm m}^{\prime}$, then we have
\begin{equation}
E_{\rm p} = \frac{1}{1+z} \frac{3 h q_{\rm e} B^{\prime}}{4 \pi m_{\rm e} c} \Gamma \gamma_{\rm m}^{\prime 2},
\label{eq:Ep}
\end{equation}
where $h$ is the Planck constant, $c$ is the speed of light, $q_{\rm e}$ and $m_{\rm e}$ are the electron charge,
and electron mass respectively, $z$ is the redshift of the burst.
$z =1$ is adopted in this work unless a specific statement.
When the jet encounters the surrounding interstellar medium, its bulk Lorentz factor $\Gamma$ will decrease.
Meanwhile, in a rapidly expanding jet, the magnetic field strength $B^{\prime}$ would decay with radius
by a general form of $B^{\prime}=B_{0}^{\prime} (R / R_0)^{-q}$, where $R$ is the radius of the expanding shell.
Since the radial component of the magnetic field decrease more rapidly than the toroidal component,
the magnetic field is soon toroidally dominated, which decreases as $R^{-1}$~\citep{Lyubarsky09}.
So $q$ is always set to be 1 in the main text~\citep{Uhm15,Geng18a},
while the effect of different $q$ (e.g., \citealt{Ronchini21}) on our results is further discussed in Appendix C.
Hence, as a function of decreasing $\Gamma$ and $B^{\prime}$,
a hard-to-soft evolution pattern for $E_\text{p}$ is naturally expected according to Equation (\ref{eq:Ep}).
Nevertheless, the real situation is probably much complicated.
The jet may be under acceleration driven by radiative pressure or the magnetic pressure gradient.
Moreover, the observed flux comes from an equal-arrival-time surface (EATS) rather than a point-like emission site~\citep{Fenimore96,Dermer04,Huang07,Geng17},
and Equation (\ref{eq:Ep}) is invalid for an integrated spectrum.
Furthermore, $\gamma_{\rm m}^{\prime}$ might vary with time due to detailed particle acceleration processes~\citep[e.g.,][]{Guo14}.
Therefore, other evolution patterns of $E_\text{p}$ are likely to arise in specific conditions.

\section{Numerical calculations}\label{sec:data}

Instead of assuming a detailed model, we try to derive some constraints on relevant parameters used in our calculations from observations.
The pulse duration $t_{\rm obs}$ could be estimated by the angular timescale of the emitting shell, i.e.,
\begin{equation}
\delta t_{\mathrm{obs}} \simeq \frac{(1 + z) R_0}{2 \Gamma^2 c},
\end{equation}
from which the corresponding $\Gamma$ could be obtained when $t_{\rm obs}$ and the initial emission radius $R_{0}$ are given.
On the other hand, the specific flux at $E_\text{p}$ in the observer frame can be estimated as~\citep{Sari98,Kumar08}
\begin{equation}
F_{\nu_{\mathrm{obs}}}=N_{\mathrm{e}} \frac{\sqrt{3} q_{\mathrm{e}}^{3} B^{\prime} \Gamma}{m_{\mathrm{e}} c^{2}} \frac{1+z}{4 \pi D_{L}^{2}},
\label{eq:Fnobs}
\end{equation}
where $N_{\mathrm{e}}$ is the total (already corrected for $4 \pi$ solid angle) number of electrons with
$\gamma_{\mathrm{e}}^{\prime}>\gamma_{\mathrm{m}}^{\prime}$ and $D_{L}$ is the luminosity distance of the burst.
Taking typical values for a GRB pulse as $E_{\text{p}} \simeq 1 \mathrm{MeV}$, $\delta t_{\text{obs}} \simeq 1 \mathrm{s}$,
and $F_{\nu_{\mathrm{obs}}} \simeq 1 \mathrm{mJy}$,
we could obtain plausible values of different sets of parameters
($\Gamma$, $\gamma_{\rm m}^{\prime}$, $B_{0}^{\prime}$, $N_{\rm inj}^{\prime}$, $R_{0}$) from Equations (\ref{eq:Ep}-\ref{eq:Fnobs}).
More details could be found in Appendix B.
Detailed parameter values are listed in Table \ref{models},
from which we derive evolving $E_{\rm p}$ (corresponding to the peak point of $\nu_{\rm obs} F_{\nu_{\rm obs}}$)
from the observed spectrum integrated over the EATS for each specific time (Appendix D).
The curvature effect on $E_{\rm p}$ and $F_{\nu_{\rm obs}}$ is thus naturally incorporated.

To explore the conditions that generate distinct evolution patterns of $E_\text{p}$,
we will show results by adding physical effects progressively in the following subsections.
A shell moving at a constant Lorenz factor with two different starting radii where the shell begins to emit photons is modeled in subsection 3.1.
Two typical dissipation radii for GRB prompt emissions, i.e., $R_0 = 10^{14}$~cm and $10^{15}$~cm, are considered.
A shell moving at an accelerating Lorenz factor is assumed and calculated in subsection 3.2.
We further investigate the case that the parameter $\gamma_{\mathrm{m}}^{\prime}$ is varying in subsection 3.3.
Moreover, two short GRBs, i.e., GRB 090227B and GRB 110529A are modeled (3.4).
Finally, the prompt optical emission is discussed in subsection 3.5.

\subsection{Constant $\Gamma$}

Here, we first assume that the expanding shell is moving at a constant Lorentz factor of $\Gamma$.
In this group of calculations, we perform seven calculations,
which are named in the form of ``$\text{CXRY}\Gamma \text{Z}$'' with the letter C representing that $\Gamma$ is constant.
X is an integer ranging from 1 to 4, which denotes cases that the cooling process of electrons is dominated by
synchrotron cooling, synchrotron self-Compton (SSC) cooling, adiabatic cooling, and SSC+adiabatic cooling at the same time in order.
Y presents the logarithmic index of the starting radius and Z shows the value of $\Gamma$.
The corresponding results, including the evolution of the electron distribution ($d N_{\rm e} / d \gamma^{\prime}_{\rm e}$),
the flux density ($F_{\nu}$), the spectral energy distribution ($\nu F_{\nu}$), lightcurves,
and the evolution of $E_{\rm p}$ for each case are exhibited in Fig. 1.

In the case of $\text{C1R14}\Gamma \text{240}$, the cooling process of electrons
is forced to be dominated by synchrotron cooling by artificially ignoring SSC cooling.
The electron distribution obeys the standard fast-cooling pattern
($dN_{\rm e}/d\gamma^{\prime}_{\rm e} \propto  {\gamma^{\prime}_{\rm e}}^{-2}$) in the early stage,
and its slope gets flatter soon since adiabatic cooling begins to dominate for low-energy electrons when $B^{\prime}$ decreases.
The corresponding synchrotron spectra and lightcurves show that $E_\text{p}$ always decreases gradually
during the rising and decaying stages of the pulse.
Thus a hard-to-soft pattern of $E_\text{p}$ arises in this case.
In $\text {C1R15}\Gamma \text {120}$, the cooling of electrons exactly trace the standard fast-cooling pattern
for the whole duration. This is owing to that adiabatic cooling rate is inversely proportional to the radius of the shell~\citep{Uhm12,Geng14}.
So, synchrotron cooling always dominates through the whole stage.
Meanwhile, $E_\text{p}$ also exhibits a hard-to-soft evolution.

SSC cooling dominates in both $\text{C}2\text{R}14 \Gamma 240$ and $\text{C}2\text{R}15 \Gamma 460$,
characterized by that the slope of the low-energy electron spectrum approaches -1~\citep{Bosnjak09,Wang09}.
Although the electron/photon spectra are harder, $E_\text{p}$ still appears a hard-to-soft evolution in both cases.
As shown in Fig. 1, for $\text{C3R14}\Gamma \text {1900}$,
the dominance of adiabatic cooling leads to an even harder low-energy electron spectrum~\citep{Geng18b}.
The evolution of $E_\text{p}$ follows an intensity-tracking pattern.
Electrons tend to gather in the low energy scope due to the inefficient cooling rate.
As a result, the slope of low-energy electron spectrum turns to be positive in one segment.
Correspondingly, the peak intensity as well as the peak frequency in $F_{\nu}$ spectrum changes slowly and slightly,
which is different from the above cases.
Equation (\ref{eq:Ep}) is now invalid and $E_\text{p}$ is related to the integral of this segment of electrons.
As an integral effect, the upwarp of the electron spectrum leads to a right shift of the peak point of $\nu F_{\nu}$ spectrum.
The electron spectrum of $dN_{e}/d\gamma^{\prime}_{\rm e} \propto {\gamma^{\prime}_{\rm e}}^{0}$ would generate a flux density of $F_{\nu} \propto {\nu}^{1/2}$.
If the slope of the low-energy spectra exceeds 0, the index of corresponding $F_{\nu}$ spectra will be greater than $1/2$.
As a result, the peak point of $\nu F_{\nu}$ spectrum will shift towards high frequencies.

Unlike $\text{C3R14}\Gamma \text{1900}$, a larger initial radius of the shell in $\text{C3R15}\Gamma \text{1900}$
results in a slower magnetic decay, and the cooling process is dominated by synchrotron cooling rather than adiabatic cooling.
In this case, $E_\text{p}$ exhibits a hard-to-soft evolution.
If SSC cooling together with adiabatic cooling governs the cooling process, the electron spectrum will quickly get harden.
As a consequence, it also shows an intensity-tracking evolution for $E_\text{p}$.
This is presented in the case of $\text{C}4\text{R}14 \Gamma 460$.

According to these results, hard-to-soft evolution for $E_\text{p}$ is normally expected in constant $\Gamma$.
The necessary condition that could generate the intensity-tracking pattern is
that the slope of the low-energy electron spectra at the early stage must be harder than -1.0.
This requires that electrons are dominated by adiabatic cooling or adiabatic+SSC cooling.

\subsection{Increasing $\Gamma$}

A bulk acceleration of $\Gamma$ may lead to the increase of $E_{\rm p}$ according to Equation (\ref{eq:Ep}).
Both radiative pressure and the magnetic pressure gradient within the jet could be responsible for the acceleration of the jet~\citep[e.g.,][]{Piran93,Drenkhahn02a,Geng21}.
Hence, we consider that the Lorentz factor of the shell is increasing gradually~\citep{Uhm18}, i.e.,
\begin{equation}
\Gamma(R)=\Gamma_{0}\left(R / R_{0}\right)^{s},
\label{Gamma}
\end{equation}
where $R$ is the radius of the shell at some time.
The acceleration index $s$ ranges from $1/3$ to 1 in different acceleration theories~\citep{Drenkhahn02a,Komissarov09}.
Three representative values, i.e., 0.35, 0.55, and 1.0 are taken in calculations.
Similar to those in the previous section, we consider two possible initial radiation radiuses of the shell.
This group of numerical results are presented in Fig. 2,
which are named in the form of ``$\text{A}\text{RY} \Gamma Z$''.
A means that the shell is accelerating,
Y is the logarithmic index of the starting radius, and Z is the initial Lorentz factor $\Gamma_{0}$ of the shell.

In $\text{A}\text{R} 14 \Gamma 200$, $s$ is set to be 0.35 and the cooling process is dominated by SSC cooling.
The low-energy electron spectrum gets hard within a very early period, while $F_{\nu}$ spectrum in the high-frequency domain does not change much over time.
$E_\text{p}$ exhibits an intensity-tracking pattern evolution, which is distinct from case $\text{C2}\text{R} 14 \Gamma 240$.
On the one hand, the increasing $\Gamma$ profits to the rise of $E_\text{p}$ analytically.
On the other hand, we find that the EATS effect is against the occurrence of the intensity-tracking pattern.
The acceleration effect weakens the EATS effect, owing to compel low latitude photons of the EATS to be more significant.
Thus the case of $\text{C2}\text{R} 14 \Gamma 240$ does not present an intensity-tracking pattern of $E_\text{p}$.
In $\text{A}\text{R} 14 \Gamma 150$, $s$ is 0.55 and SSC cooling is also dominant.
The slope of the low-energy electron spectrum tends to be -1 early and exceeds -1 at the late stage.
The evolution curve of $E_\text{p}$ also shows an intensity-tracking pattern.
In $\text{A}\text{R} 14 \Gamma 80 $, $s$ is 1.0.
The electron distribution and the flux density are similar to those in $\text{A}\text{R}14 \Gamma 150$.
$E_\text{p}$ presents an intensity-tracking evolution again.

As for the case of $\text{A}\text{R}15 \Gamma 230$, $s$ is 1.0,
a hard-to-soft evolution for $E_{\rm p}$ is shown,
which is a natural result from the dominance of synchrotron cooling and the EATS effect.
Owing to a larger initial radius, the photon energy density is reduced significantly
so that it is hard to make SSC cooling dominate the cooling process at an early time.
As a result, the low-energy electron spectra follow the standard cooling except for the last stage.
Moreover, the moving distance of the shell during $\delta t_{\rm obs}$ is only the magnitude of $R_0$, which implies a strong EATS effect on $F_{\nu}$.

In this subsection, the significance of SSC cooling, as well as the bulk acceleration effect, are discussed.
SSC cooling could not produce an intensity-tracking pattern of $E_\text{p}$ alone, due to the impediment of the EATS effect.
However, an accelerating shell working together with a dominant SSC cooling could generate an intensity-tracking pattern of $E_\text{p}$
for small initial radius of $10^{14}$~cm.
For a larger initial radius of $10^{15}$~cm, only a hard-to-soft pattern is resulted even for
an accelerating shell because of the absence of SSC cooling dominance.

\subsection{A variable source function}

The particle-in-cell simulations for relativistic magnetic reconnections indicate that
the accelerated particle spectrum is evolving with time~\citep[e.g.,][]{Guo14}.
Here, we take a variable source function into account.
Specifically, a broken power-law profile of $\gamma_{\mathrm{m}}^{\prime}(R)$ \citep{Uhm18} is assumed,
\begin{equation}
\gamma_{\mathrm{m}}^{\prime}(R)={\gamma_{\mathrm{m}}^{0}} \times\left\{\begin{array}{ll}
\left(R / R_{\rm a}\right)^{g} & \text { if } R \leqslant R_{\rm a}, \\
\left(R / R_{\rm a}\right)^{-g} & \text { if } R > R_{\rm a},
\end{array}\right.
\label{gm}
\end{equation}
where $R_{\rm a}$ is a radius that particle acceleration efficiency begins to decrease.
In the real situation, the turnover time for particle acceleration efficiency
is probably not identical with $t_{\rm off}$.
Thus, $R_{\rm a}$ is set to be smaller than $R_{\rm off}$ which corresponds to $t_{\rm off}$.
Similarly, four cases are included in this group of numerical results (see Table \ref{models}),
and the corresponding figures are exhibited in Fig. \ref{fig:3}.
The cases are named in the form of ``$\text{AGX}\text{RY} \Gamma Z$'',
where G means that the evolution of $\gamma_{\mathrm{m}}^{\prime}(R)$ is taken into account,
X is an integer ranging from 1 to 2, denoting the bulk acceleration effect is considered or not, respectively.
The meanings of Y and Z are similar to those in subsection 3.2.

In $\text{AG}1\text{R}14 \Gamma 300$, SSC cooling is dominant at the early stage and the slope of the low-energy electron spectra tends to -1,
which is the same as the case of $\text{AG}2\text{R}14 \Gamma 150$.
In $\text{AG}1\text{R}14 \Gamma 300$, we first consider the evolution of $\gamma_{\text{m}}^{\prime}$ only but ignore the bulk acceleration effect.
It presents a chromatic intensity-tracking pattern for $E_{\rm p}$, which means the peak time of $E_\text{p}$ and lightcurves is mismatched.
When the bulk acceleration effect is also included in $\text{A}2\text{R} 14 \Gamma 150$,
it shows a chromatic intensity-tracking too.
This chromatic intensity-tracking pattern for $E_{\rm p}$ has not been discovered from observations.
Different from the above intensity-tracking patterns, the chromatic intensity-tracking pattern is induced by
the effect of a variable source function which reflects the intrinsic particle acceleration.
An intensity-tracking pattern is expected for $E_\text{p}$ when $R_{\rm a}$ is set to be the same value as $R_{\rm off}$.
If this new pattern is confirmed from detailed data analyses in the future,
it would be evidence for the varying $\gamma_{\text{m}}^{\prime}$ from particle acceleration.

In results of $\text{AG}1\text{R}15 \Gamma 250$, owing to the large initial radius,
synchrotron cooling merely dominates the whole cooling process and SSC cooling emerges at the last stage.
Also, a large initial radius makes the EATS effect strongly affect the observed flux density
due to little movement of the shell relative to the initial radius (also see Appendix B).
Consequently, the role of a varying $\gamma_{\text{m}}^{\prime}$ is suppressed by the synchrotron cooling dominance and the EATS effect.
Hence it is understandable that $E_\text{p}$ shows a hard-to-soft evolution pattern.
While for $\text{AG}2\text{R}15 \Gamma 230$, both the bulk acceleration effect and the varying $\gamma_{\text{m}}^{\prime}$ are considered.
The EATS effect is weakened so that $E_\text{p}$ presents a chromatic intensity-tracking pattern again at the very early stage.

\subsection{Application to GRB data}

To test the applicability of our calculations, we perform theoretical modeling of two representative short GRBs, i.e.,
GRB 110529A (intensity-tracking) and GRB 090227B (hard-to-soft) respectively.
The data used from {\it Fermi}-GBM\footnote{https://heasarc.gsfc.nasa.gov/W3Browse/fermi/fermigbrst.html}
is analyzed with the standard Bayesian approach and the time bin selection method to obtain the time-resolved spectra (see details in~\citealt{Li19a}).
Because of the huge computational cost in each calculation
and potential strong degeneracy between the parameters,
a strict fitting procedure with the Bayesian method is not carried out in the current work.
Instead, we change the physical parameters involved manually through some trials,
until a visual good match between the numerical results and the data is reached.
The fitting parameters are listed in Table~\ref{fitp}.
As is shown in Fig.~\ref{fig:fitting}, the evolution of $E_{\rm p}$ and flux for these two GRBs could be roughly reproduced.

In GRB 110529A, the shell is under accelerating and SSC cooling is dominant,
making a rapid rise of the lightcurve and an intensity-tracking pattern of $E_{\rm p}$.
In GRB 090227B, the shell is moving at a constant Lorentz factor
and the cooling process is dominated by synchrotron cooling, resulting in a hard-to-soft pattern of $E_{\rm p}$.
Note that there is a distinct sub-pulse within 0.1-0.15 s as shown in photon counts of GRB 090227B,
which leads to the significant deviation of theoretical $E_{\rm p}$ from the observed ones during this period.

\begin{table*}
\caption{The fitting parameters.}
\label{Table:results}
\centering
\begin{tabular}{c c c c c c c c c}
\hline\hline
GRB & $\Gamma$ & $\gamma^{\prime}_{m}$ & $B^{\prime}_{0}$    & $N^{\prime}_{\text {inj}}$  & $R_{0}$  & $s$  \\

      &          & $(10^{4})$    &($10^{2}$ G) &$(10^{47} \text {s}^{-1}$) & (cm)\\
\hline
110529A  & 300  & 3.0   & 0.6     & 1.4  &  $10^{14}$    &$0.35$\\
090227B  & 240  & 5.0   & 2     & 4.0  &  $10^{14}$          \\
\hline
\end{tabular}
\tablefoot{We assume a redshift of $z=1$ for both GRBs during fitting.}
\label{fitp}
\end{table*}

\subsection{The prompt optical emission}

Relevant researches indicate that we need more probes to distinguish possible radiation mechanisms,
except for the evolution patterns of $E_\text{p}$ and the low-energy spectral indices.
Therefore, in this subsection, we calculate the prompt optical emission of the above fifteen distinct cases.
The representative results are exhibited in Fig. \ref{optical},
including $\text{C}1\text{R}15 \Gamma 120$, $\text{C}2\text{R} 15 \Gamma 460$, and $\text{A}\text{R}14 \Gamma 150$.

As shown in Fig. \ref{optical}, the flux of the prompt optical emission is higher than the prompt X-ray emission but within only one magnitude.
A clear spectral lag and a shallower decay for the optical emission are present.
The detection sensitivity of the Ground-based Wide Angle Camera (GWAC) and future Wide Field Survey Telescope (WFST) of China
is roughly 15 and 20 apparent magnitudes in the $R$ band within an exposure time of one second.
Accordingly, prompt synchrotron optical emissions from a burst within $\sim$ 1 Gpc are detectable by GWAC ($\sim$ 10 Gpc by WFST).
Future observations on the prompt optical emission will help to examine the synchrotron radiation scenarios.

\section{Conclusions and discussion} \label{sec:comare}

In this article, we numerically trace the evolution of electrons and the corresponding
synchrotron emissions in the framework of synchrotron radiation, to explore possible evolution patterns of $E_\text{p}$.
The effects of the bulk acceleration of the emitting shell, the varying $\gamma_{\text{m}}^{\prime}$,
and the EATS are analyzed by performing groups of calculations.
In addition, the prompt optical emission is calculated. The following conclusions are drawn.

1. For a shell moving at a constant Lorentz factor, the intensity-tracking pattern of $E_\text{p}$
could emerge when electrons are dominated by adiabatic cooling or adiabatic+SSC cooling.
For the case of pure dominance by adiabatic cooling,
a bulk Lorentz factor $\Gamma$ of $> 1000$ for the shell
and a small emitting radius of $R_0 \le 10^{14}$~cm is required.
Otherwise, a hard-to-soft pattern is normally expected.

2. For a shell under bulk acceleration, a dominant SSC cooling could generate an intensity-tracking pattern of $E_\text{p}$
for a small initial radius of $10^{14}$~cm.
However, for a larger initial radius of $10^{15}$~cm, only a hard-to-soft pattern is resulted
even for an accelerating shell owing to the synchrotron cooling dominance and the EATS effect.

3. When the role of the intrinsic evolution of $\gamma_{\mathrm{m}}^{\prime}$ is taken into account,
a chromatic intensity-tracking for $E_\text{p}$ may occur.
Also, the rapid decay of $B^{\prime}$ ($q \ge 1.5$), accompanied with a dominant adiabatic cooling,
would generate a chromatic intensity-tracking for $E_{\rm p}$ (Appendix C).

4. The flux density of the prompt optical emission is slightly higher than the prompt X-ray emission,
and is detectable for bursts within $\sim$ 1Gpc by current wide-field telescopes like GWAC,
and $\sim$ 10Gpc by future WFST of China.

Recently, \citet{Ronchini21} show a relation between the spectral index and the flux
by investigating the X-ray tails of bright GRB pulses, which is incompatible with the long-standing scenario
which invokes the delayed arrival of photons from high-latitude parts of the jet.
It is suggested that a dominating adiabatic cooling process, accompanied by the decay of the magnetic field
could explain this relation after testing several possible cooling mechanisms.
This study supports that adiabatic cooling could play a significant role in the GRB prompt phase.

These distinct evolution patterns, i.e., hard-to-soft, intensity-tracking and chromatic intensity-tracking,
reflect different physical processes in the GRB prompt emission.
In general, the intensity-tracking pattern of $E_{\rm p}$ indicates that the bulk Lorentz factor is initially very large or it is under acceleration,
and the emitting radius should be relatively small.
The evolution of $\gamma_{\mathrm{m}}^{\prime}$ from the particle acceleration process could generate a chromatic intensity-tracking pattern.
Further studies could help to test our conclusions.

Although only the intensity-tracking pattern of a single short pulse of the GRB is given in our work,
the result is applicable for a long-duration burst that consists of several pulses.
Owing to the proportional relation between $F_{\nu_{\mathrm{obs}}}$ and $E_\text{p}$ in time-resolved pulses observationally,
it is naturally expected that a long duration burst presents the same evolution pattern of $E_\text{p}$ as the single pulse.

In the current work, the source function $Q (\gamma_{\rm e}^{\prime},t^{\prime})$ is set the form of power-law,
i.e., $Q (\gamma_{\rm e}^{\prime},t^{\prime}) = Q_0 (t^{\prime}) (\gamma_{\rm e}^{\prime} / \gamma_{\rm m}^{\prime})^{-p}$,
which should be derived from detailed modeling of the particle acceleration process in our future works.
On the other hand, we have assumed that the shell is homogeneous in latitude and its thickness could be ignored.
Further considerations on the structure of the shell may be specified in the future.

Except for studies on the evolution pattern of $E_\text{p}$ within the synchrotron radiation scenarios,
its interpretation within the framework of the photosphere radiation is urgently invoked.
Also, better indicators (like $E_{\rm p}$, the curvature of the spectrum, etc.) to identify the underlying radiation mechanism are now under discussion.
For example, the characteristics of the low energy band (i.e., less than 1 keV) for GRB prompt emissions
may show distinct properties within different physical scenarios~\citep{Oganesyan18,Toffano21}.
Therefore, theoretical and observational efforts are encouraged in the research of the prompt optical emission.

\begin{acknowledgements}
We appreciate the anonymous referee for constructive suggestions.
We acknowledge the use of the public data from the {\it Fermi} data archives.
We also would like to thank Xue-Feng Wu for the helpful discussion and Liang Li for providing the GRB data.
This work is supported by National SKA Program of China No. 2020SKA0120300,
by the National Natural Science Foundation of China (Grant Nos. 11903019, 11873030, 11833003, 12041306, U1938201, 11535005),
and by the Strategic Priority Research Program of the Chinese Academy of Sciences
(``multi-waveband Gravitational-Wave Universe'', Grant No. XDB23040000).
\end{acknowledgements}

\bibliographystyle{aa}
\bibliography{reference}

\begin{figure*}
\centering
\begin{adjustwidth}{-1.2cm}{-1.0cm}
    \subfloat{\includegraphics[width=10.5cm,height=10.6cm]{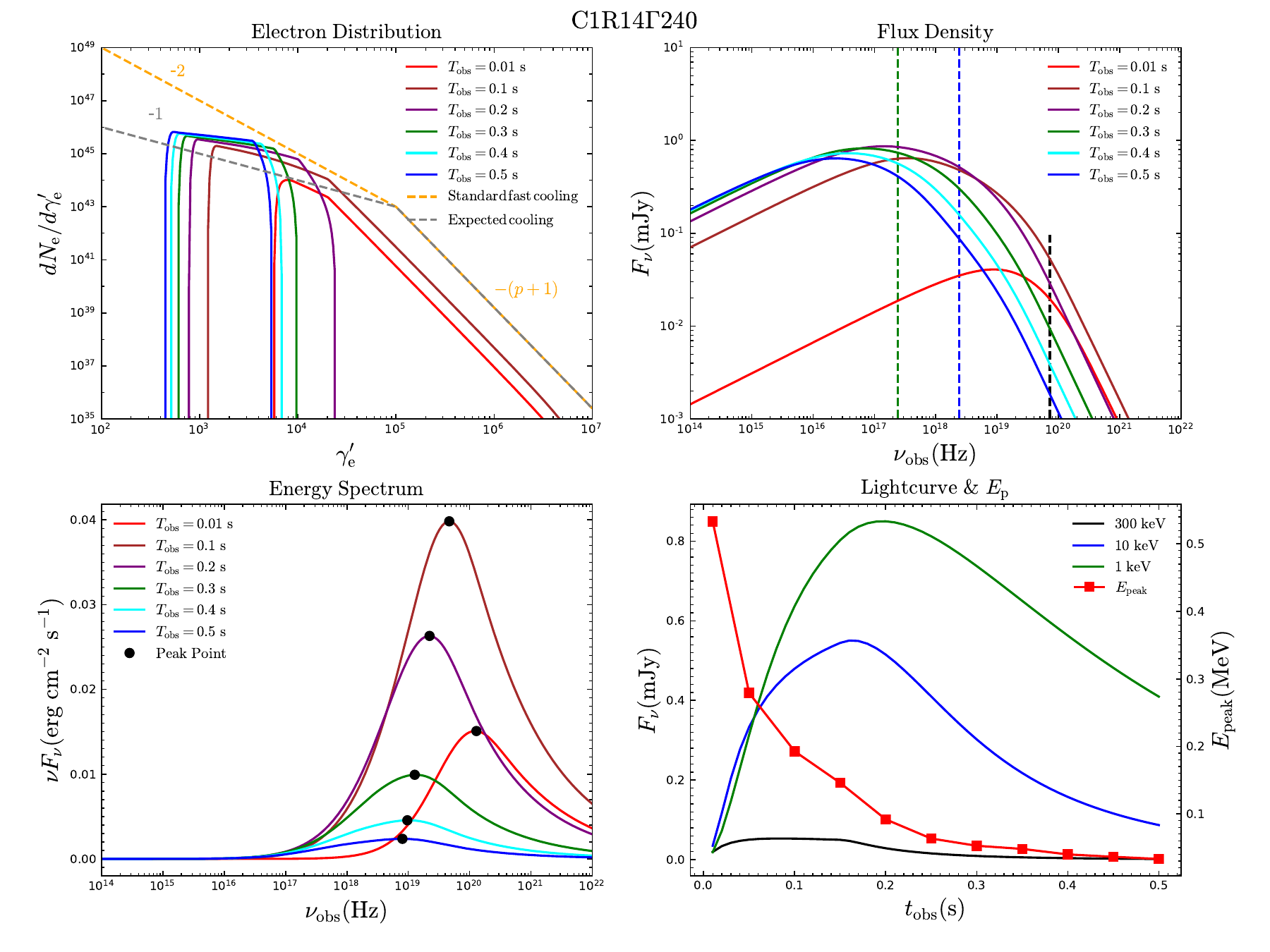}}
    \subfloat{\includegraphics[width=10.5cm,height=10.6cm]{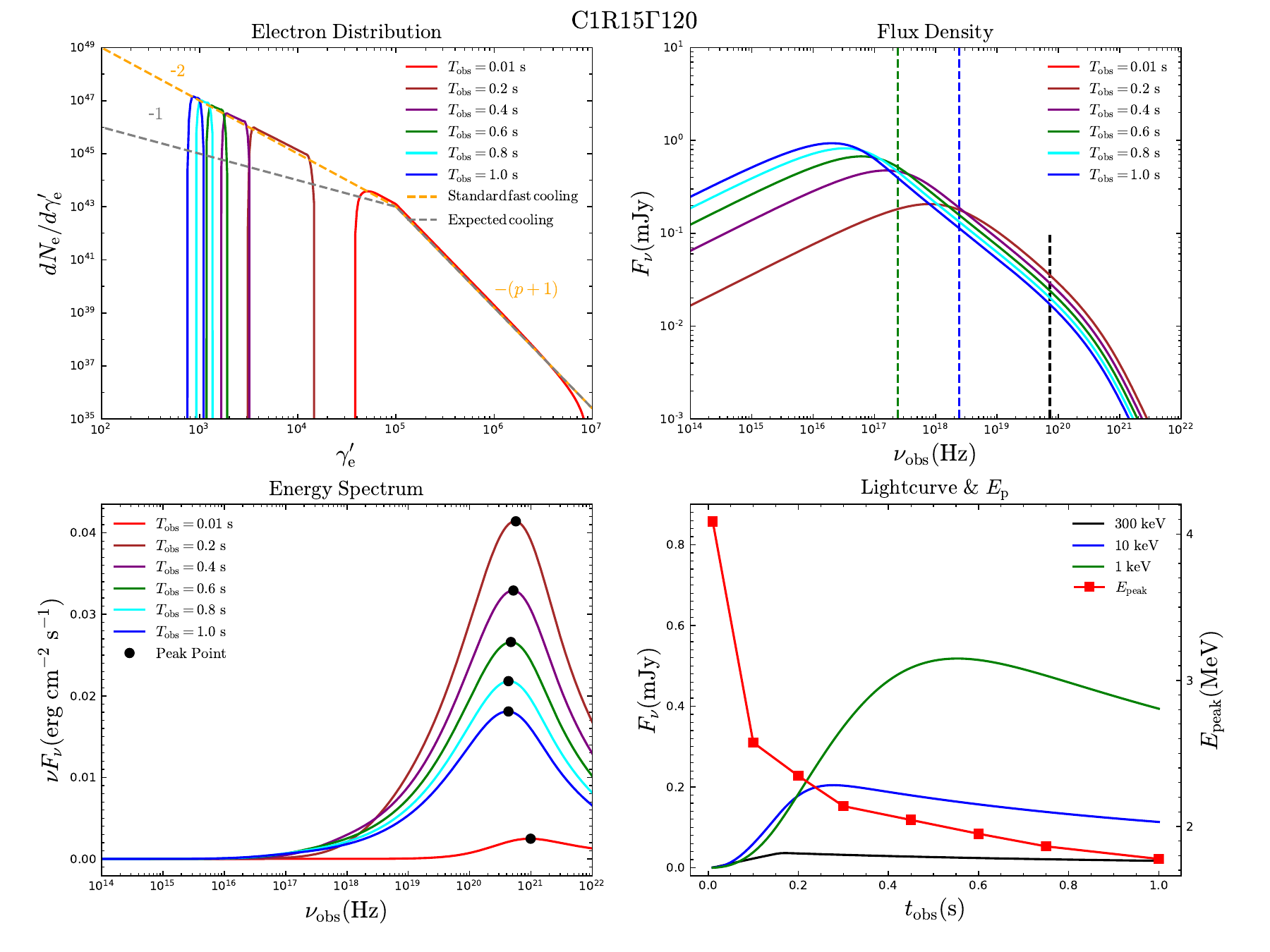}} \\
    \subfloat{\includegraphics[width=10.5cm,height=10.6cm]{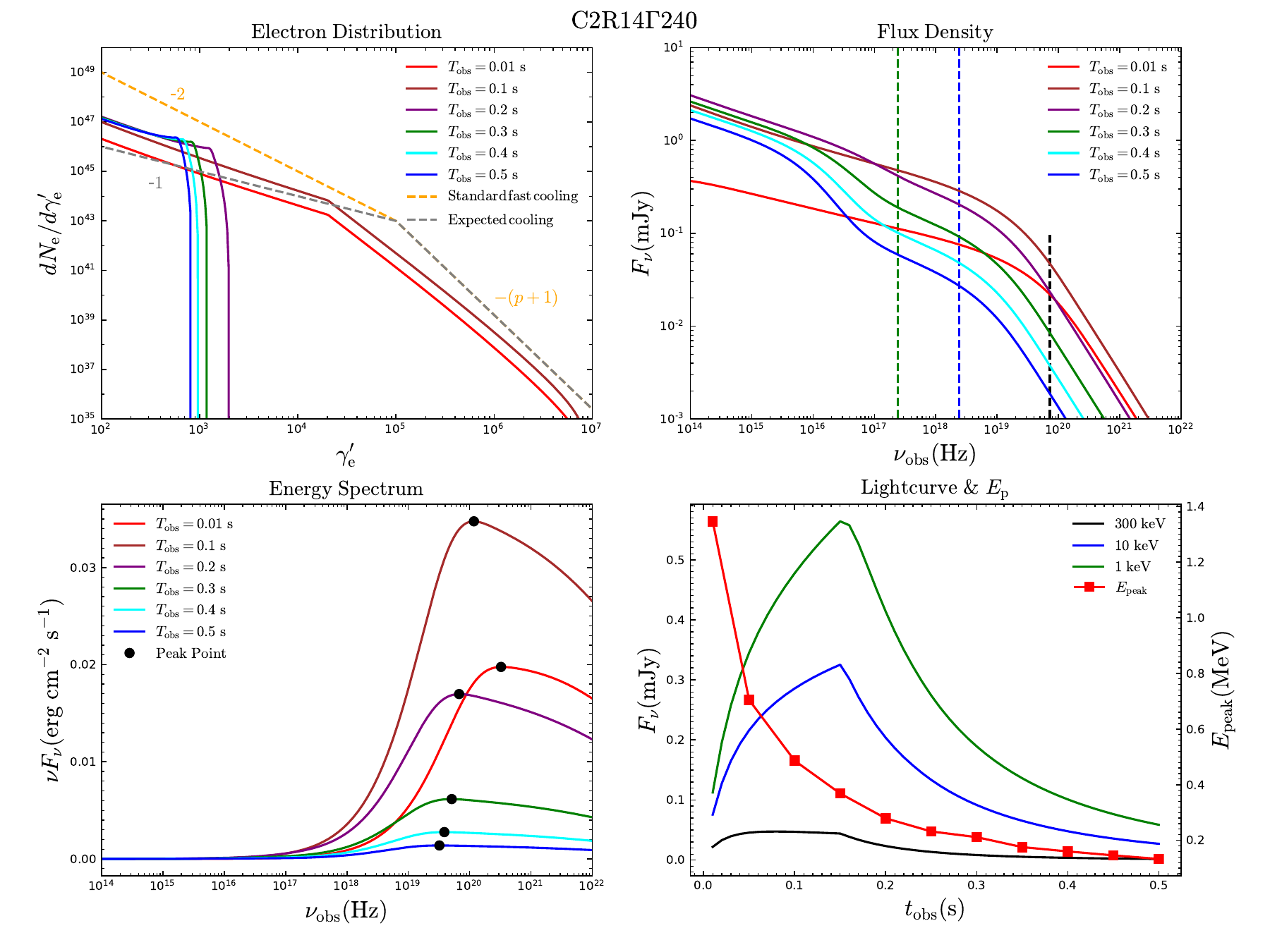}}
    \subfloat{\includegraphics[width=10.5cm,height=10.6cm]{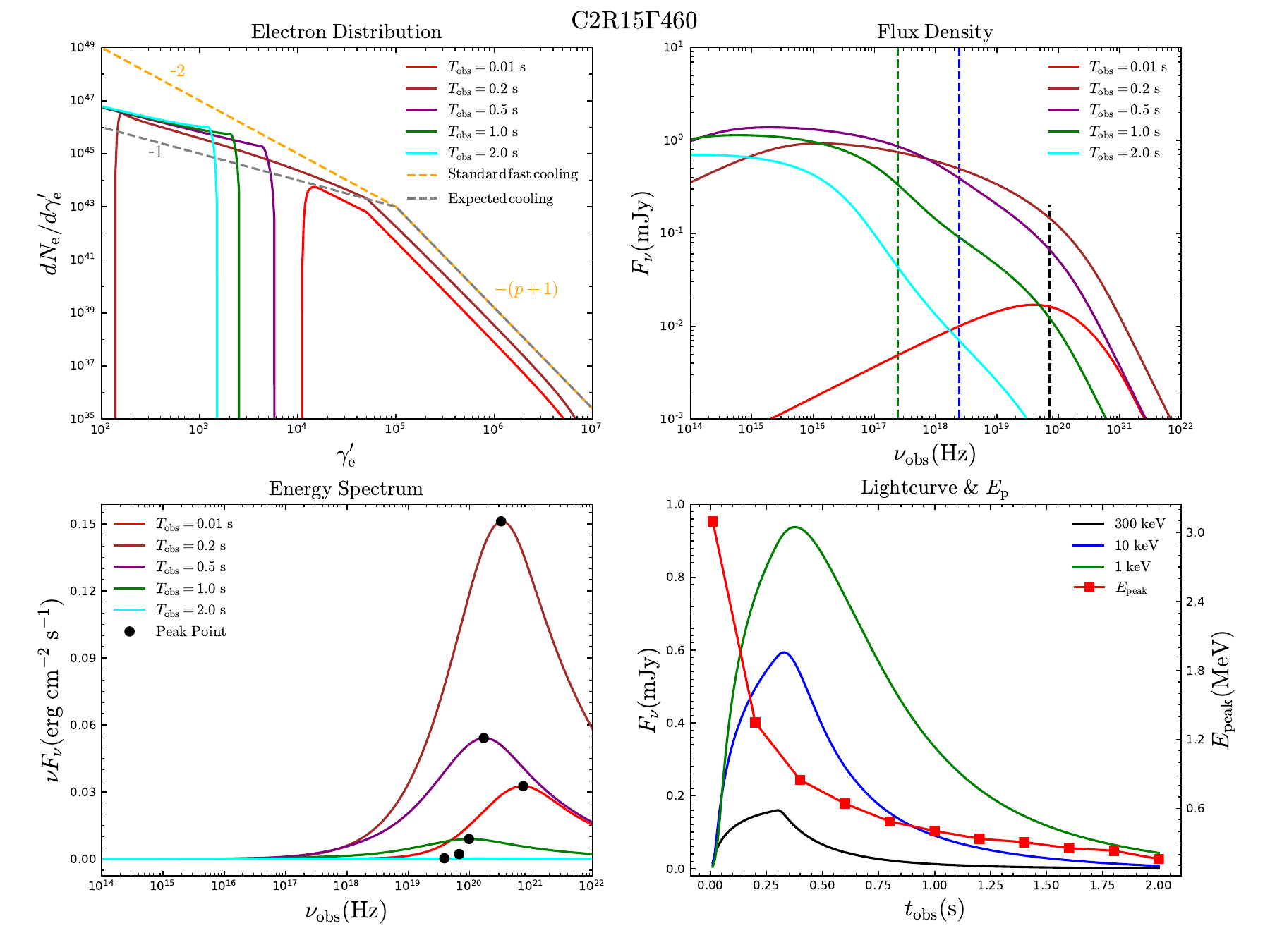}}
    \caption{Numerical results for seven cases in the first group shown in Table \ref{models}. The diagram of each case consists of 4 panels:
    the electron distribution $dN_{e}/d\gamma'_{e}$ (upper left),
    the corresponding synchrotron flux density $F_{\nu}$ from electrons (upper right),
    the spectral energy distribution $\nu F_{\nu}$ (lower left),
    and lightcurves at three bands $\&$ the evolution of $E_\text{p}$ (lower right).
    Notably, in the electron distribution, orange dashed lines are the standard fast-cooling pattern, i.e.,
    $d N_{\mathrm{e}}/d\gamma_{\mathrm{e}}^{\prime}\propto \gamma_{\mathrm{e}}^{\prime-2}$,
    and gray dashed lines present the expected cooling pattern, i.e., $d N_{\mathrm{e}} / d \gamma_{\mathrm{e}}^{\prime} \propto \gamma_{\mathrm{e}}^{\prime-1}$.
    In the flux density, three vertical lines are given corresponding to three different frequencies of 300 keV, 10 keV, and 1 keV respectively.
    The peak energy $E_\text{p}$ is derived from the peak point of $\nu F_{\nu}$.}
\end{adjustwidth}
\end{figure*}

\begin{figure*}
\ContinuedFloat
\centering
\begin{adjustwidth}{-1.2cm}{-1.0cm}
\centering
    \subfloat{\includegraphics[width=10.5cm,height=10.6cm]{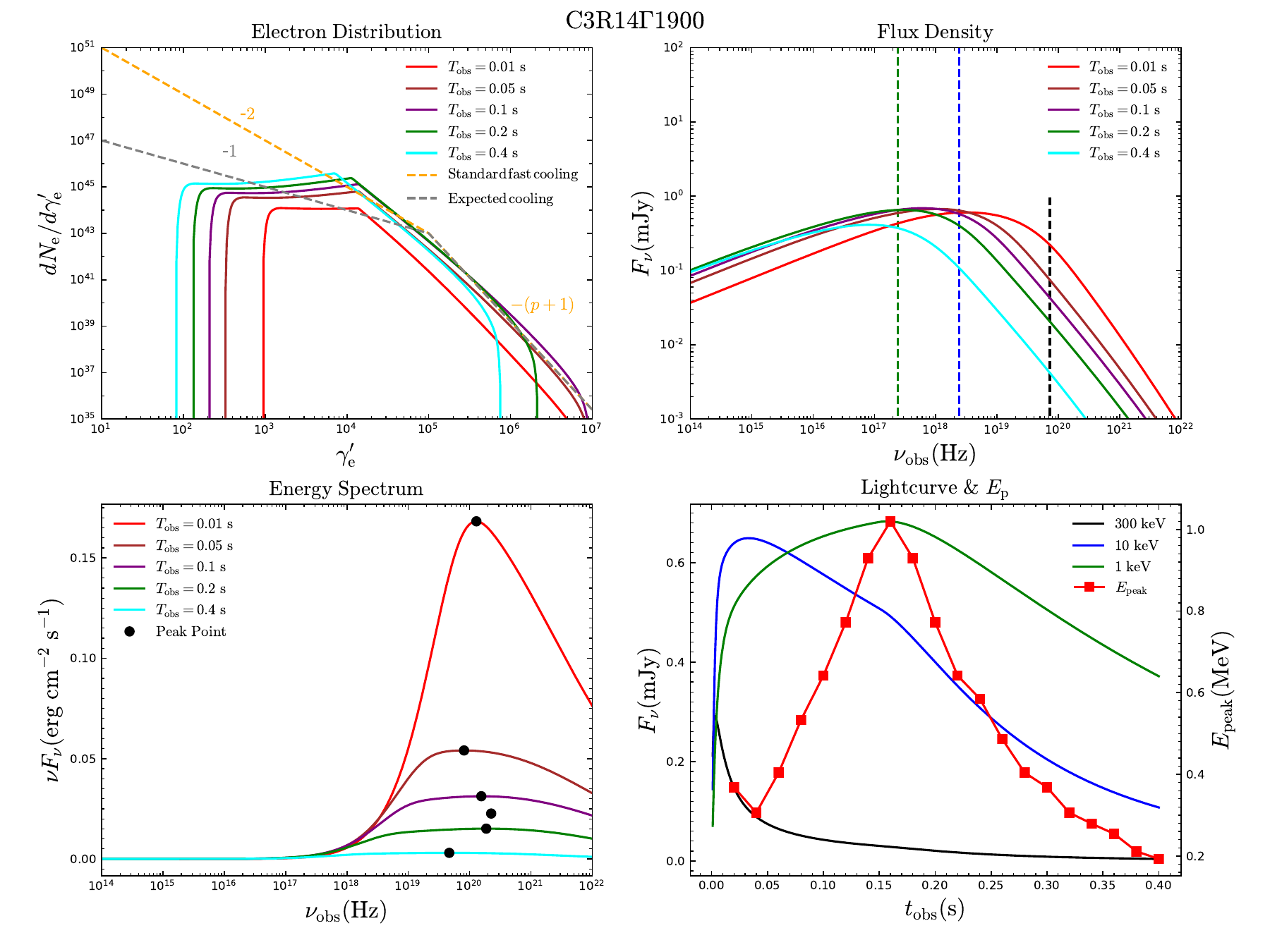}}
    \subfloat{\includegraphics[width=10.5cm,height=10.6cm]{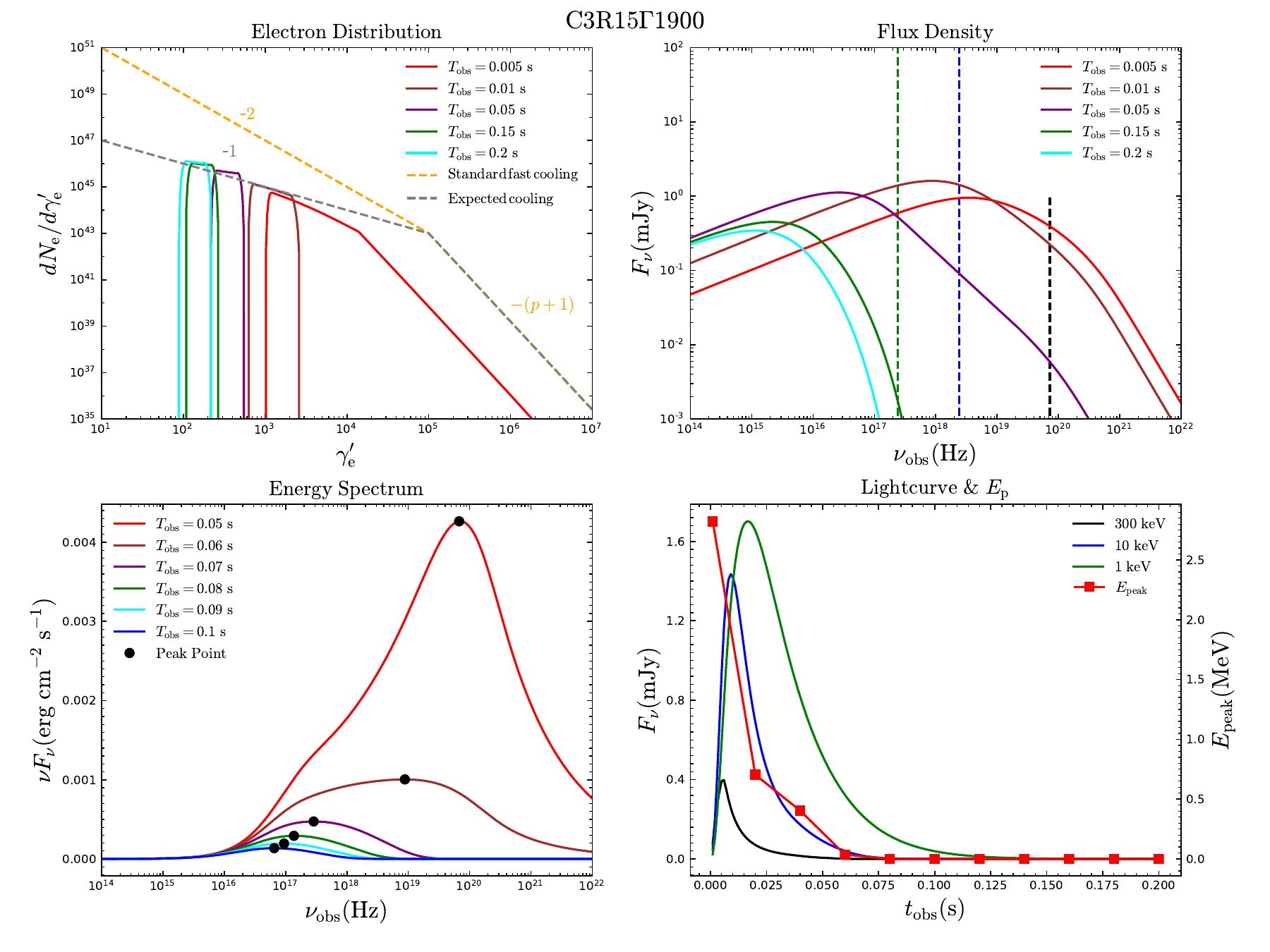}} \\
    \subfloat{\includegraphics[width=10.5cm,height=10.6cm]{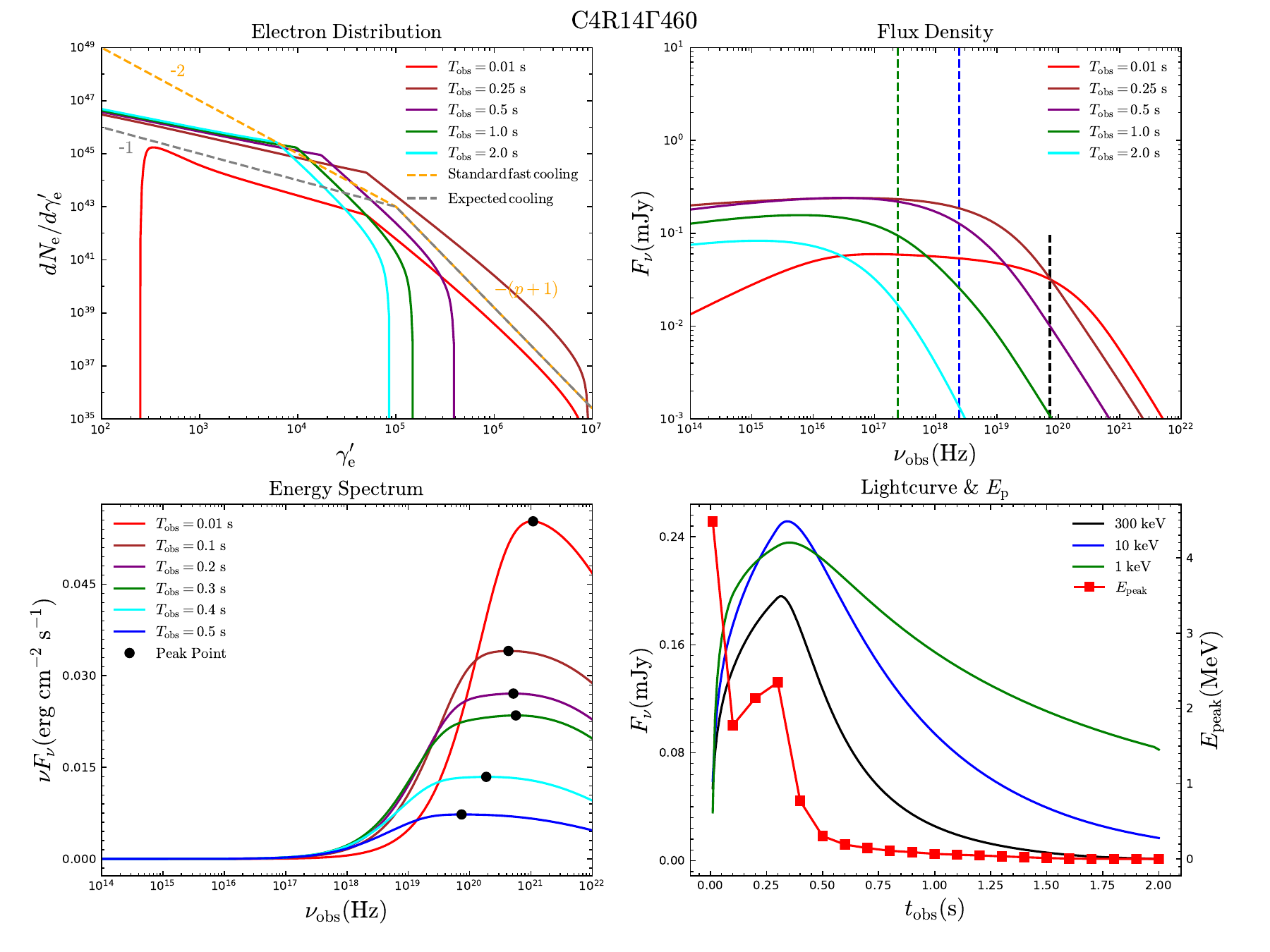}}
    \caption{(continued.)}
\end{adjustwidth}
\label{groupa}
\end{figure*}

\begin{figure*}
\centering
\begin{adjustwidth}{-1.2cm}{-1.0cm}
    \subfloat{\includegraphics[width=10.5cm,height=10.6cm]{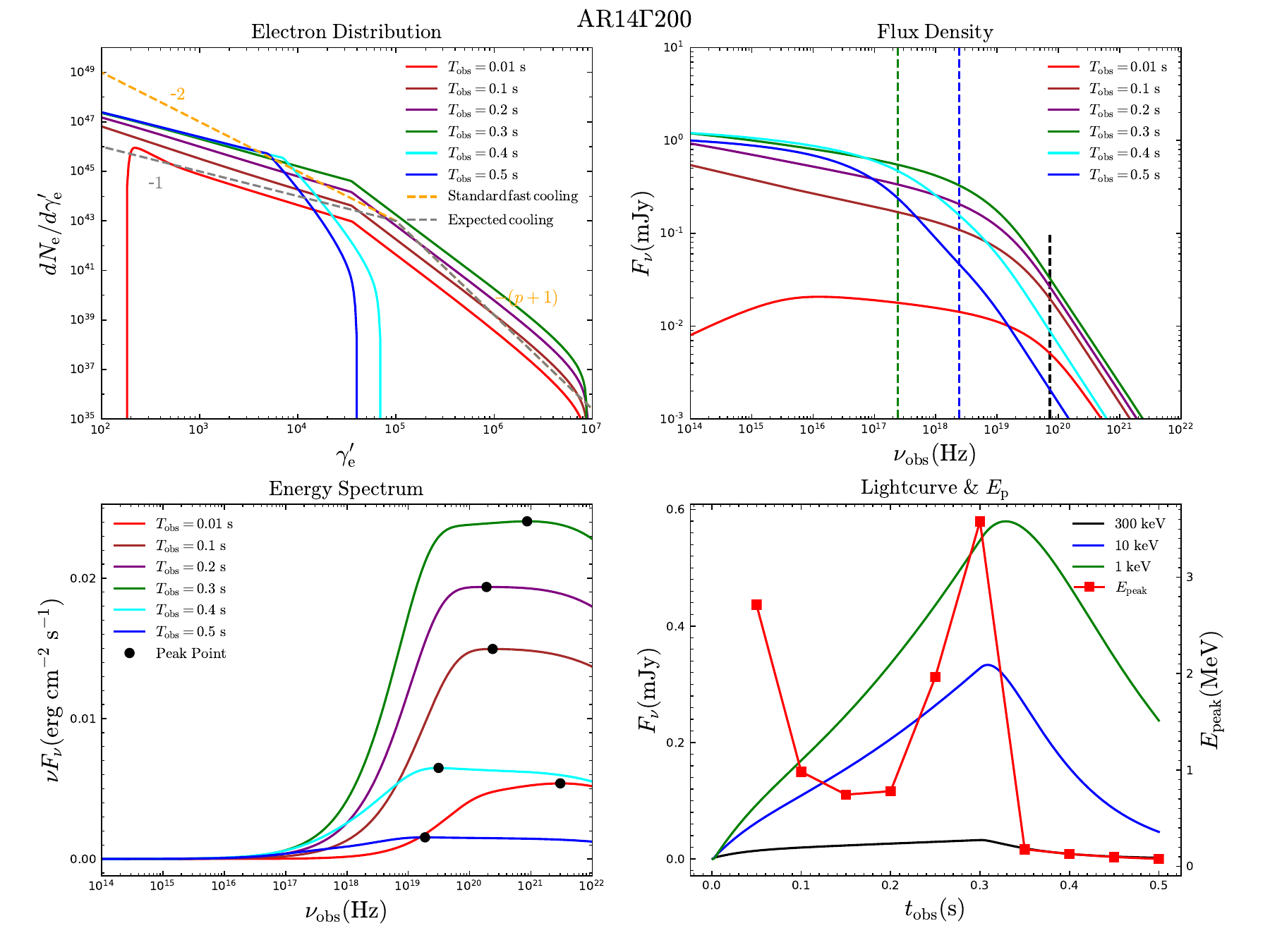}}
    \subfloat{\includegraphics[width=10.5cm,height=10.6cm]{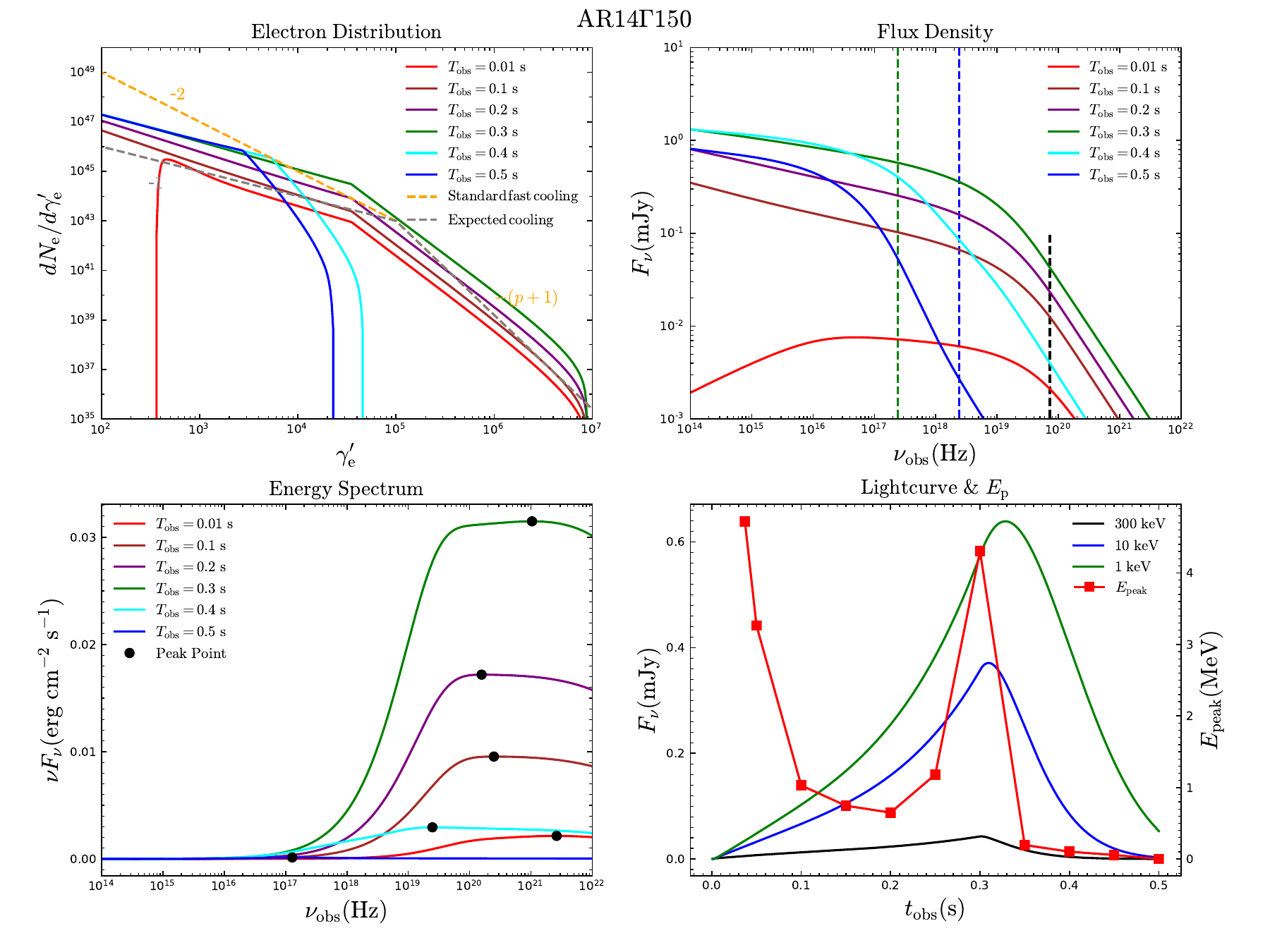}} \\
    \subfloat{\includegraphics[width=10.5cm,height=10.6cm]{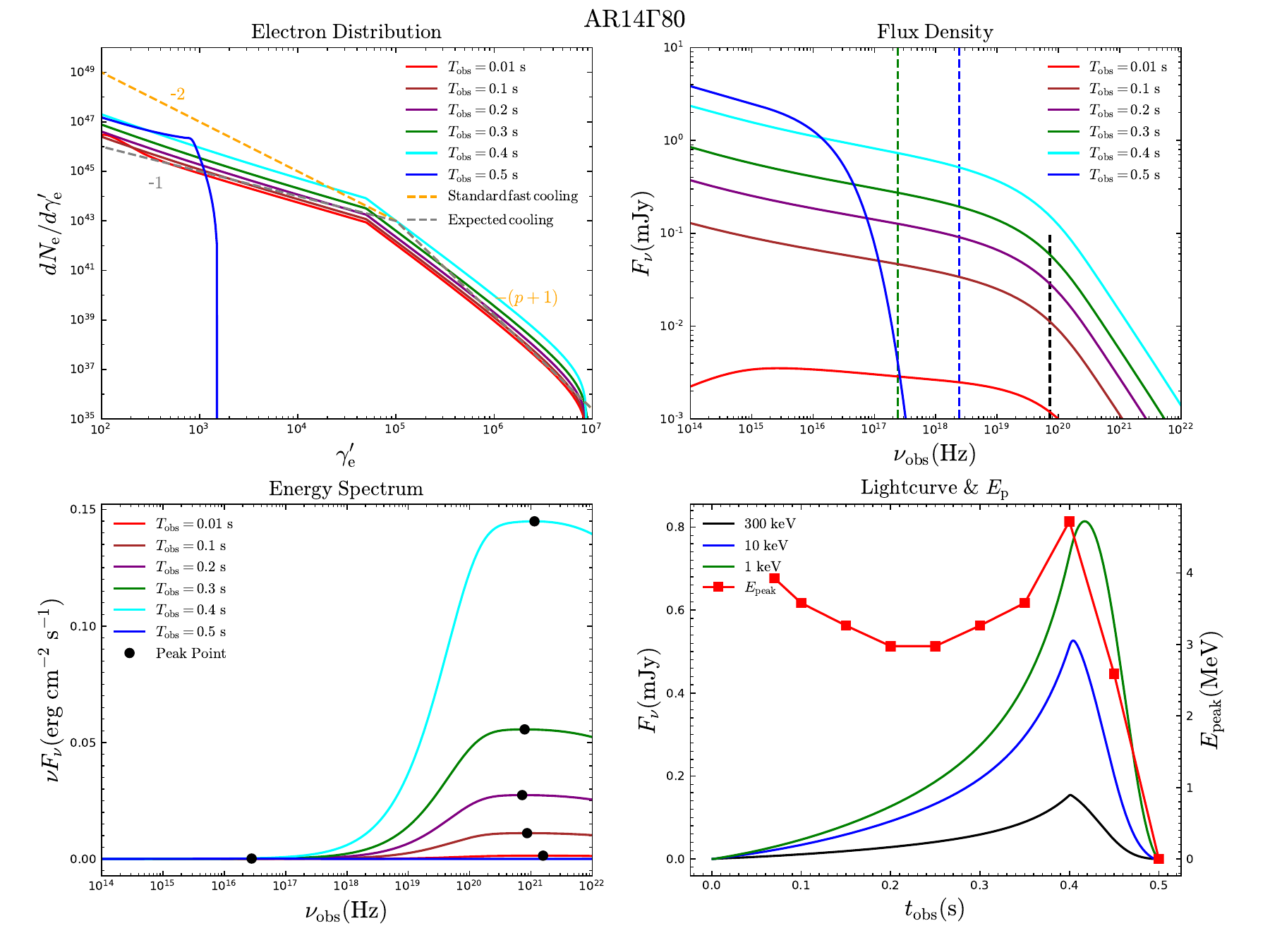}}
    \subfloat{\includegraphics[width=10.5cm,height=10.6cm]{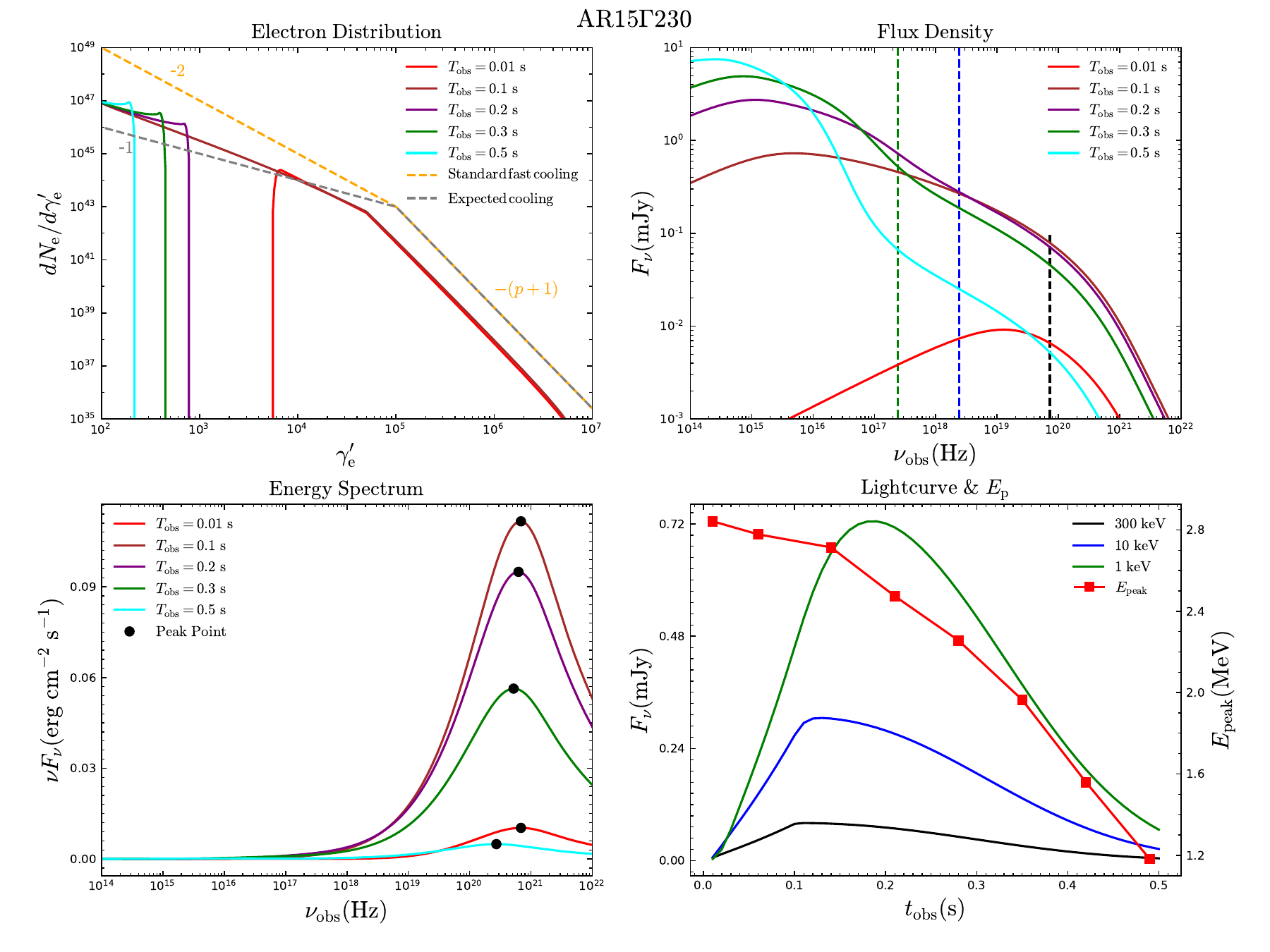}}
    \caption{Numerical results for four cases in the second group shown in Table \ref{models}.}
        \label{fig:testing1}
\end{adjustwidth}
\label{fig:2}
\end{figure*}

\begin{figure*}
\centering
\begin{adjustwidth}{-1.2cm}{0.0cm}
    \subfloat{\includegraphics[width=10.5cm,height=10.6cm]{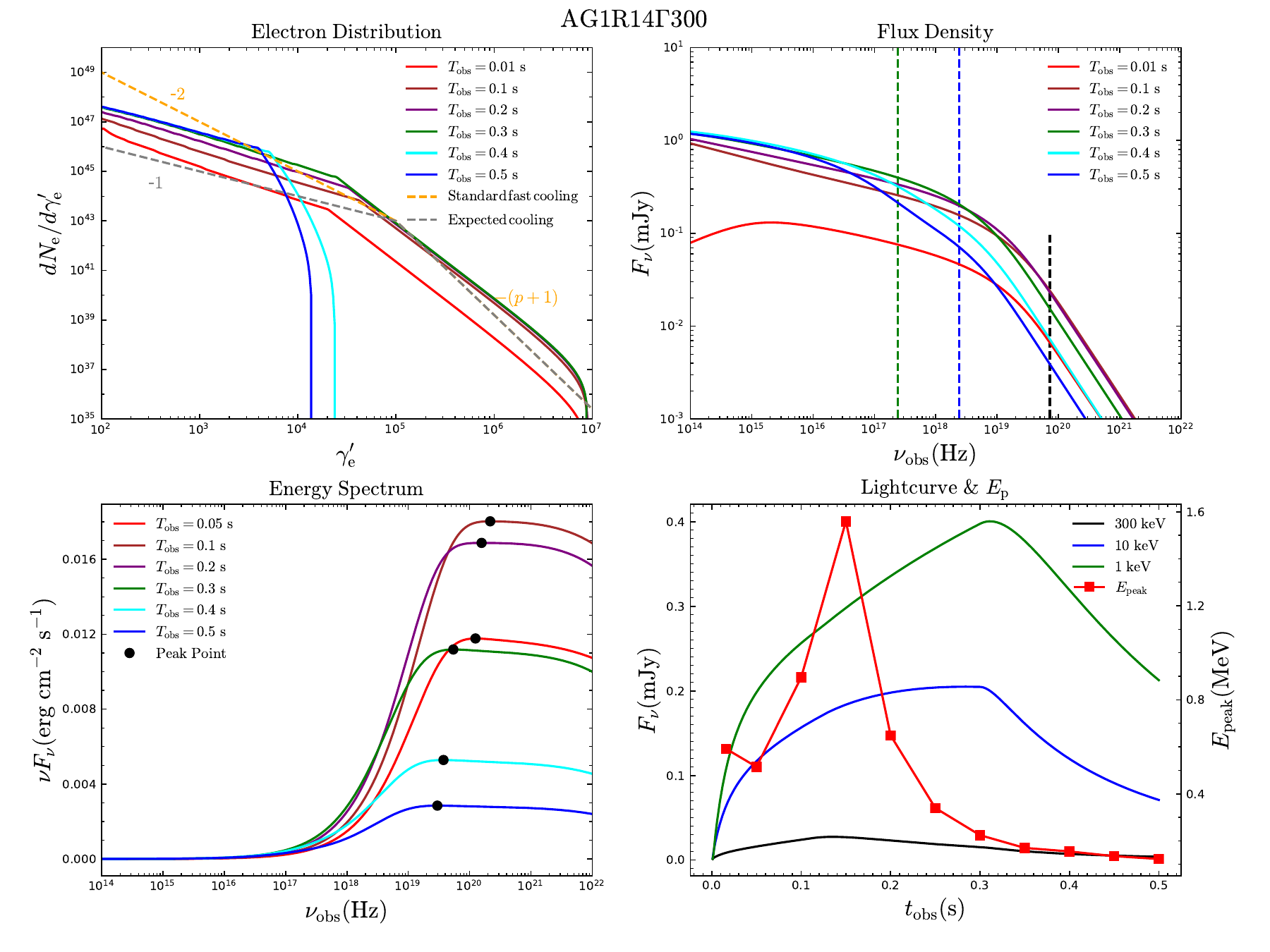}}
    \subfloat{\includegraphics[width=10.5cm,height=10.6cm]{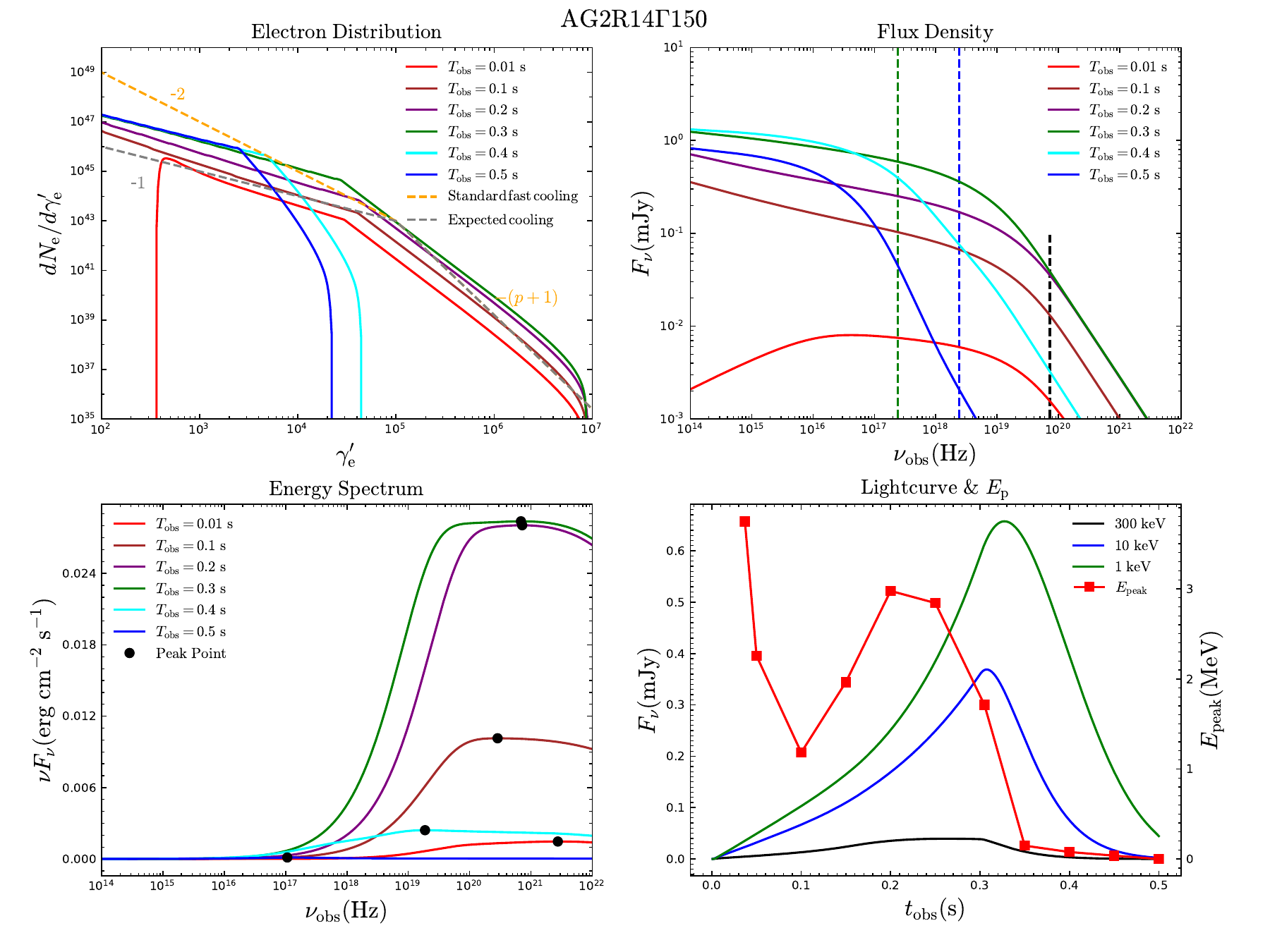}} \\
    \subfloat{\includegraphics[width=10.5cm,height=10.6cm]{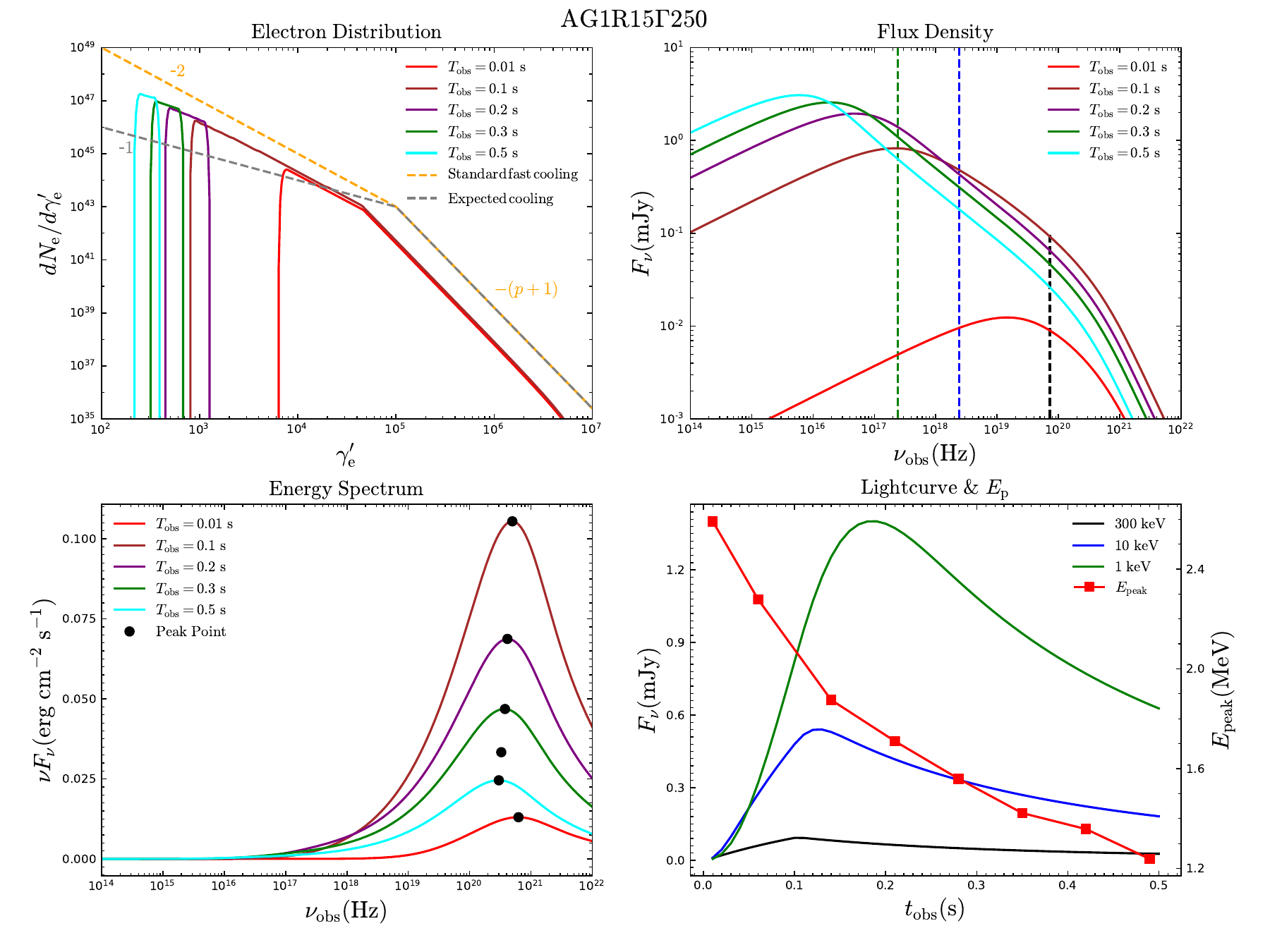}}
    \subfloat{\includegraphics[width=10.5cm,height=10.6cm]{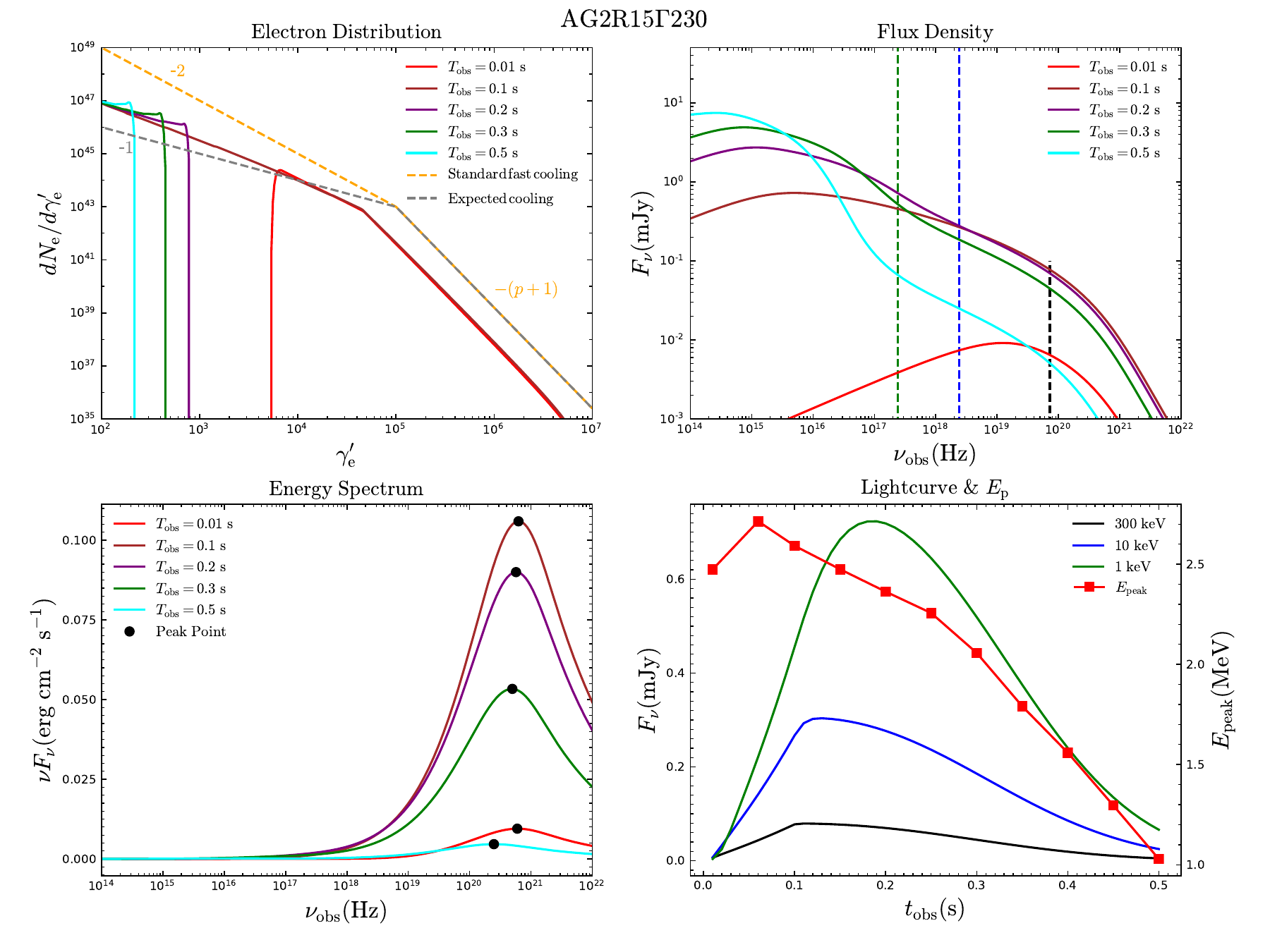}}
\end{adjustwidth}
\caption{Numerical results for four cases in the third group shown in Table \ref{models}.}
\label{fig:3}
\end{figure*}
\clearpage

\begin{figure*}
\centering
\begin{adjustwidth}{-1.2cm}{-1.0cm}
    \subfloat{\includegraphics[width=10.5cm,height=10.6cm]{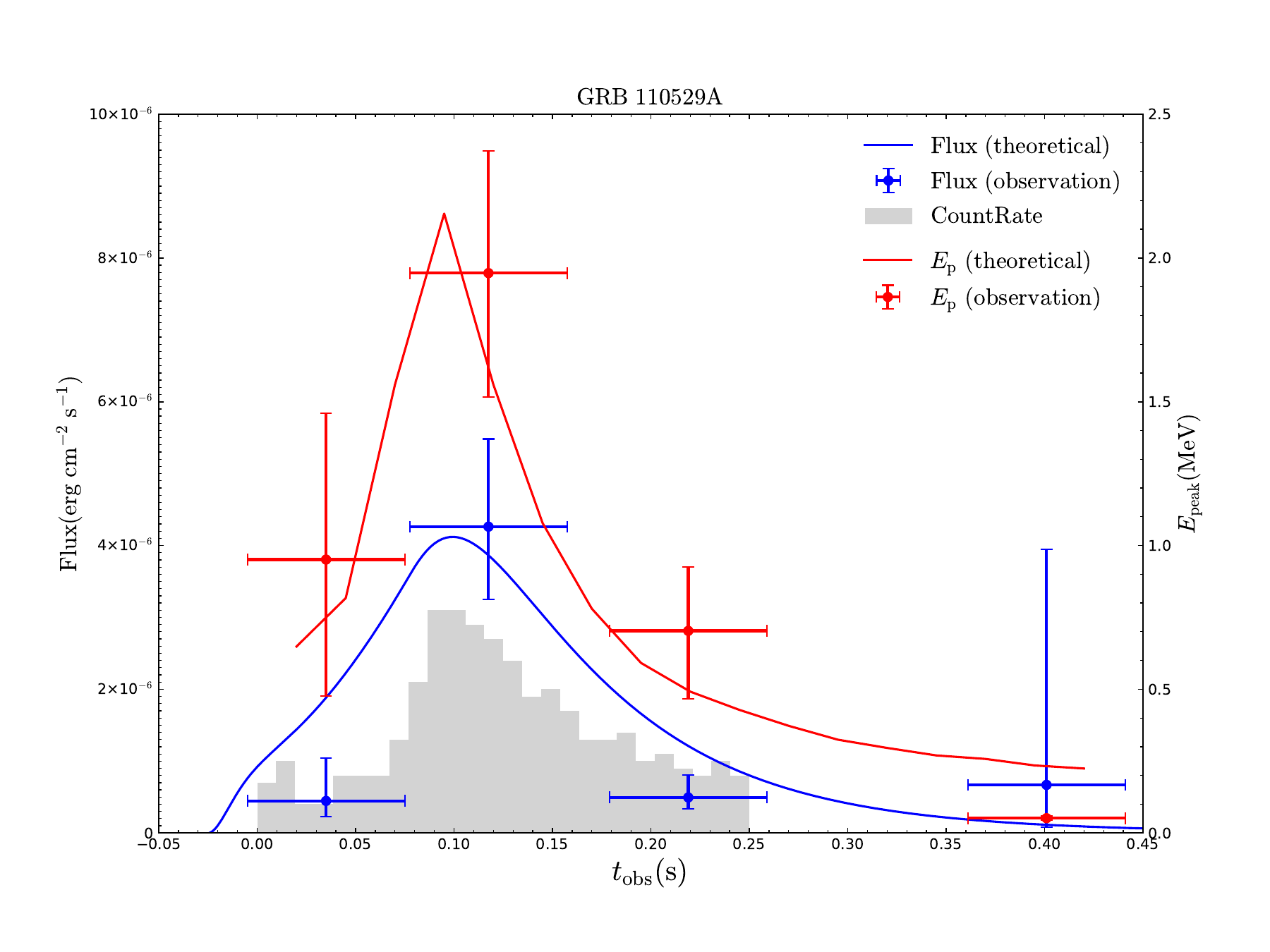}}
    \subfloat{\includegraphics[width=10.5cm,height=10.6cm]{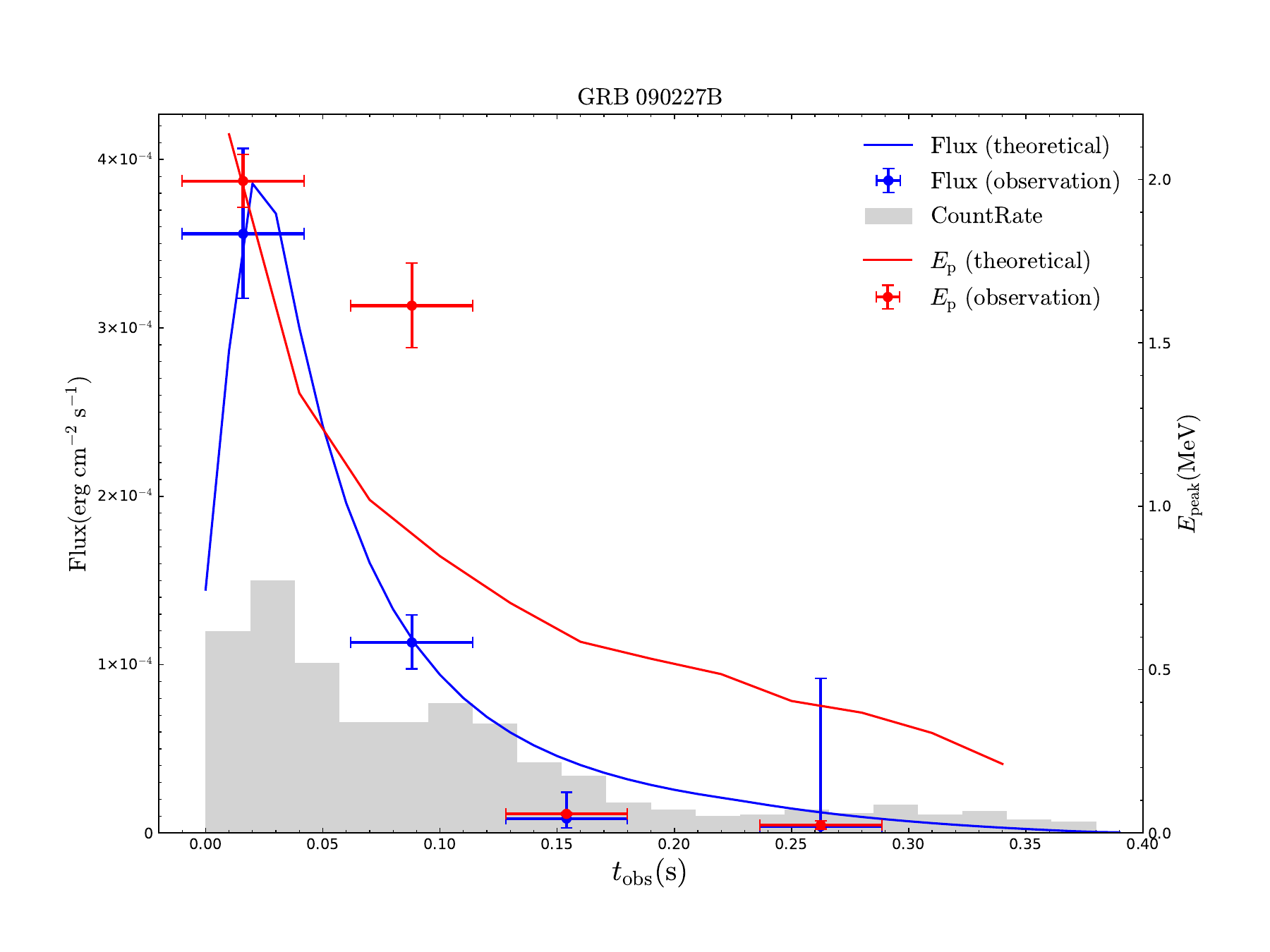}}
\end{adjustwidth}
\caption{Theoretical modeling of lightcurves and $E_{\rm p}$ evolutions for GRB 110529A and GRB 090227B.
In each diagram, numerical results are shown with blue (for flux) and red (for $E_{\rm p}$) lines,
while the observational data are presented by blue (for flux) and red (for $E_{\rm p}$) dots.
The grey histogram shows the photon-count lightcurve in a scaled way.}
\label{fig:fitting}
\end{figure*}

\begin{figure*}
\centering
    \subfloat{\includegraphics[width=6.5cm,height=8.66cm]{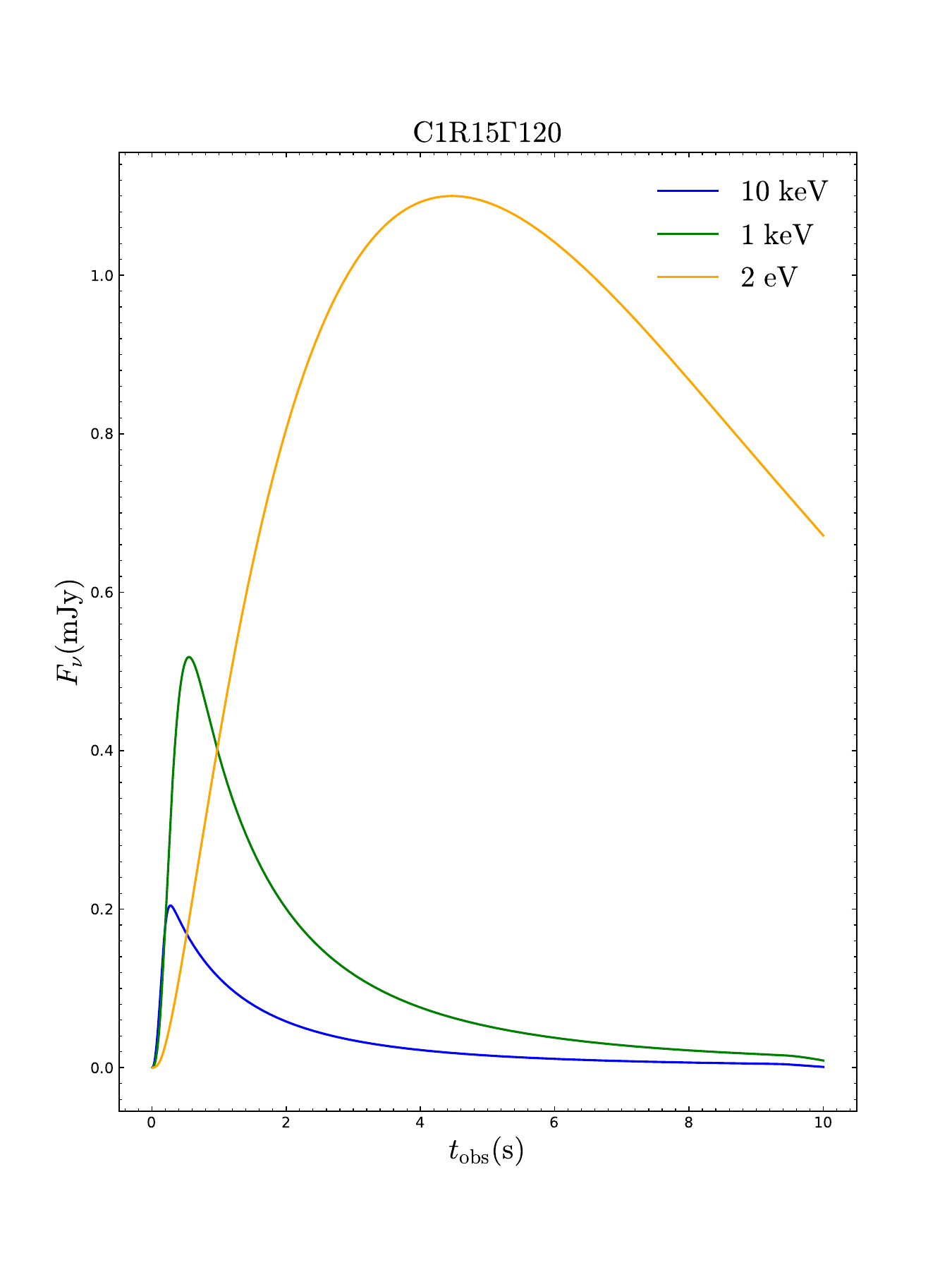}}
    \subfloat{\includegraphics[width=6.5cm,height=8.66cm]{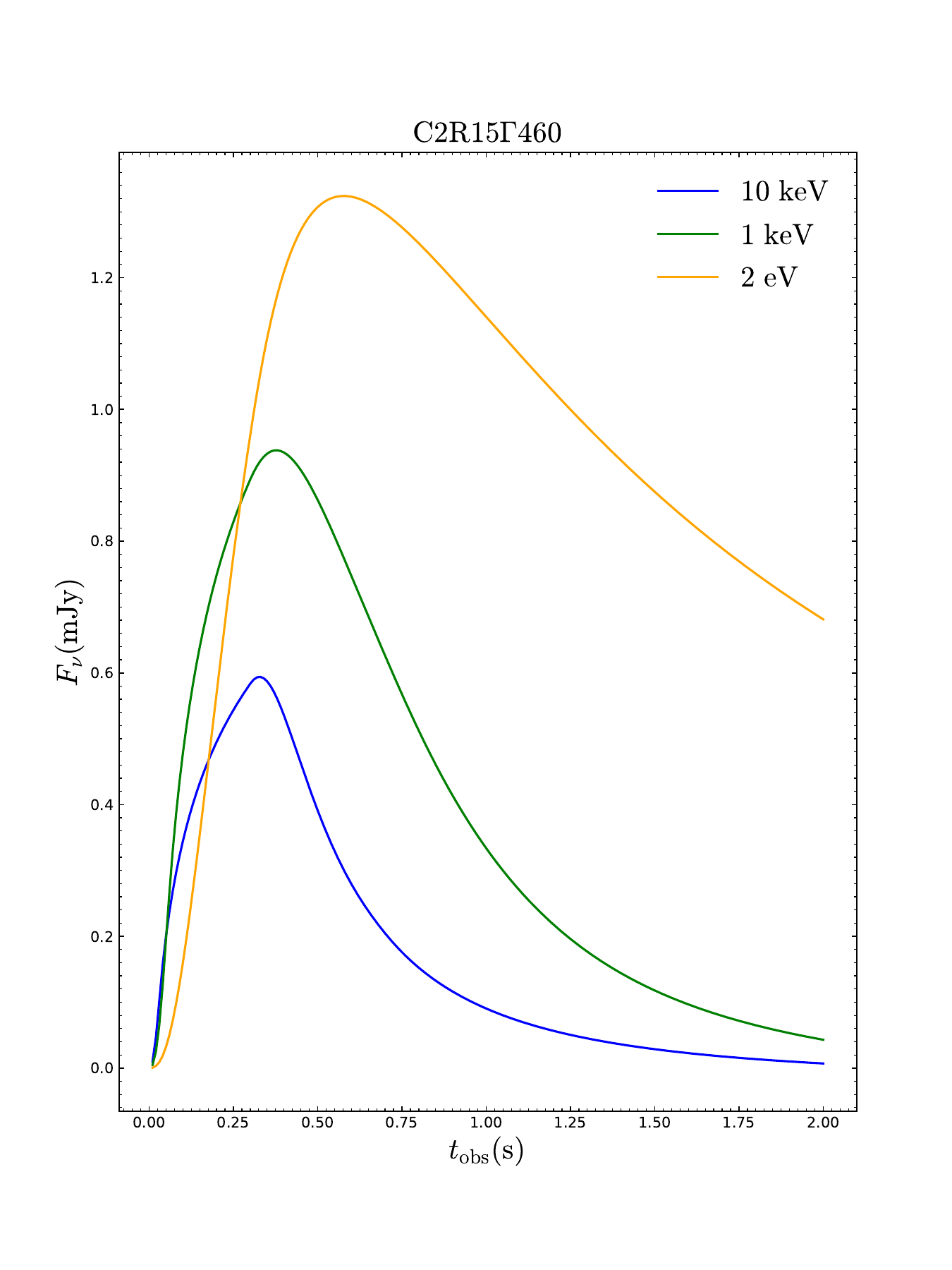}}
    \subfloat{\includegraphics[width=6.5cm,height=8.66cm]{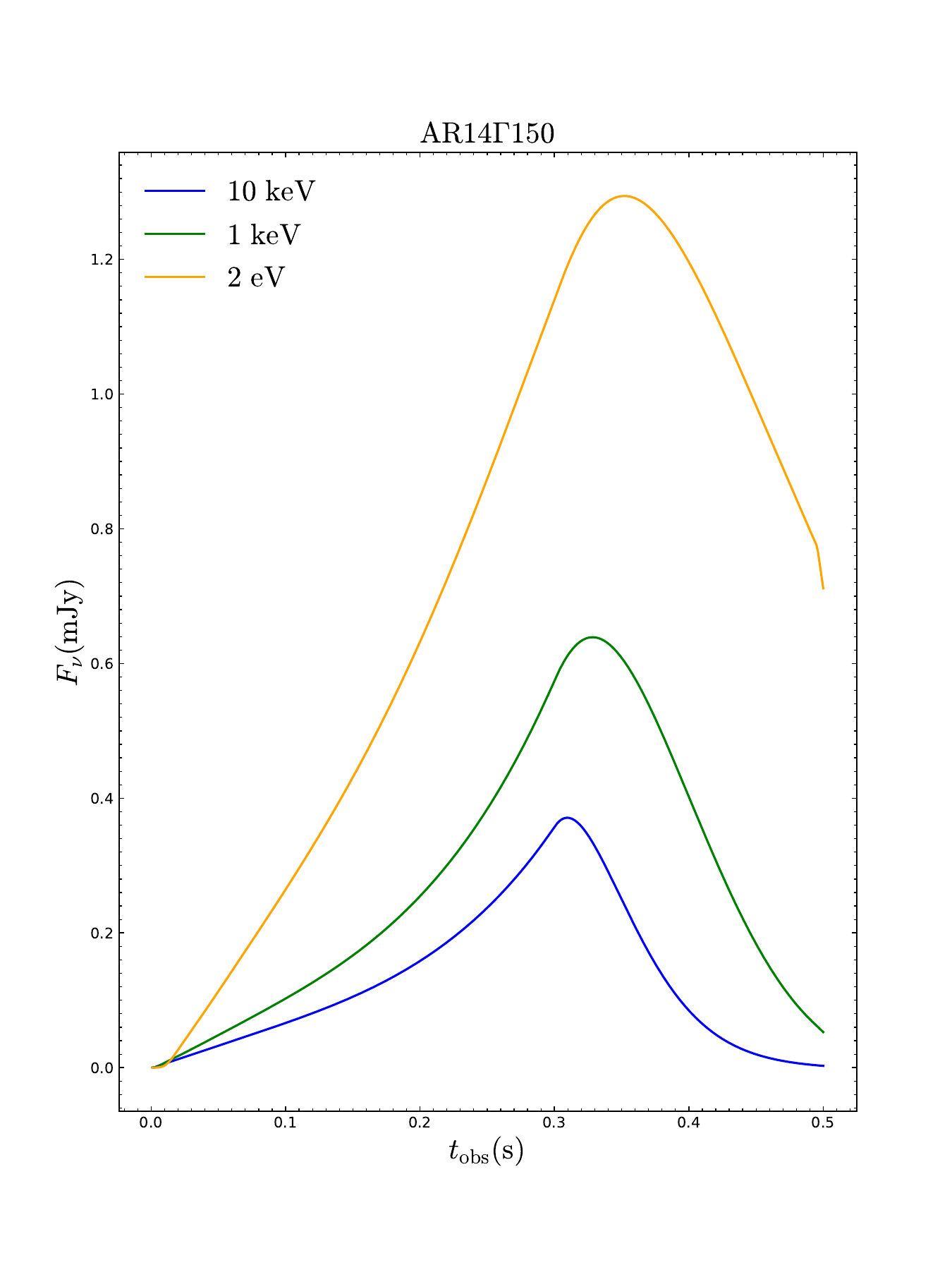}}
\caption{Representative results of the prompt optical emission from three cases in Table \ref{models}.
There are three lightcurves in different energy band in the diagram of each case, including 2 eV, 1 keV, and 10 keV.}
\label{optical}
\end{figure*}

\begin{figure*}
\centering
\begin{adjustwidth}{-1.2cm}{-1.0cm}
    \subfloat{\includegraphics[width=10.5cm,height=10.6cm]{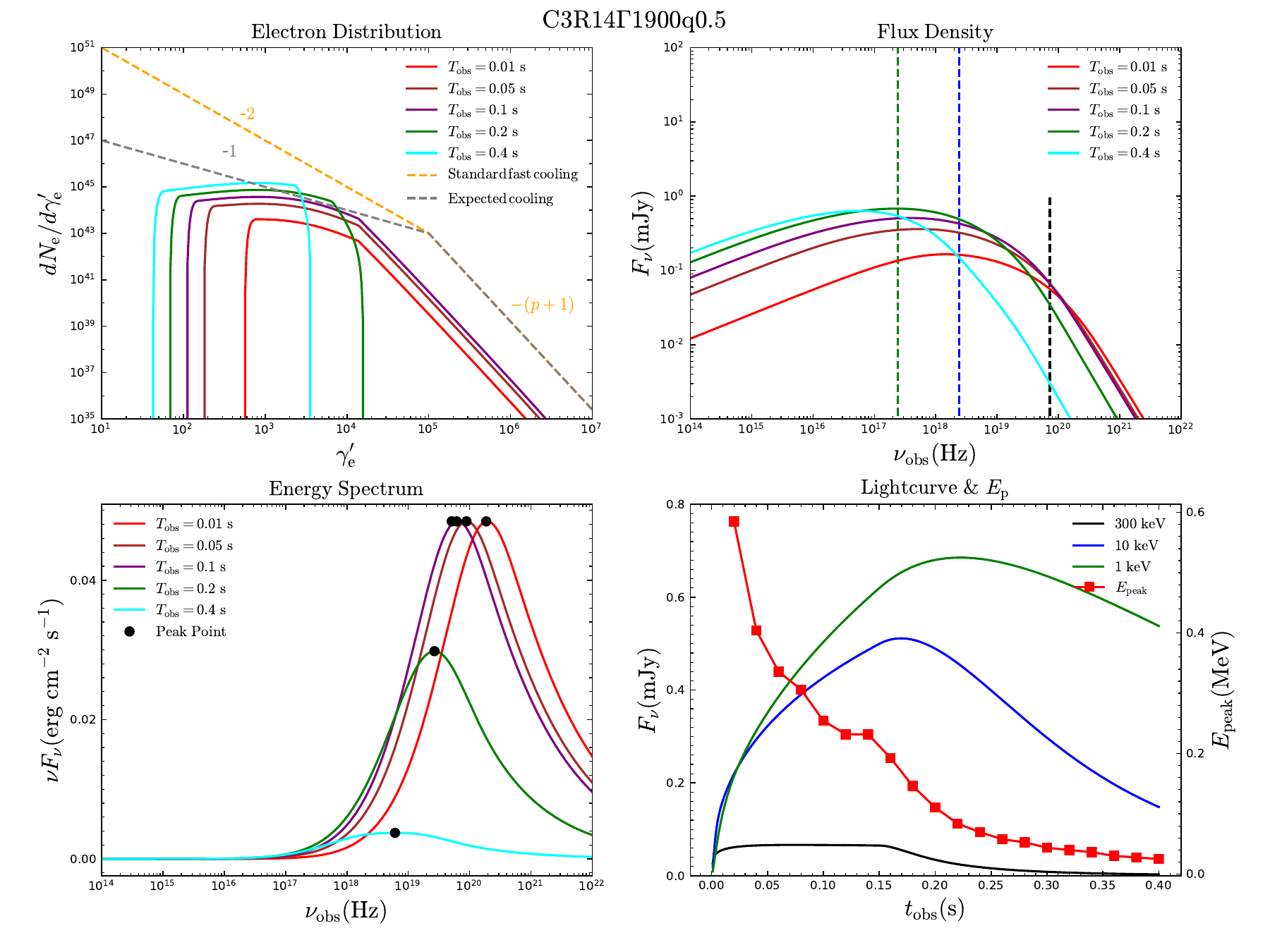}}
    \subfloat{\includegraphics[width=10.5cm,height=10.6cm]{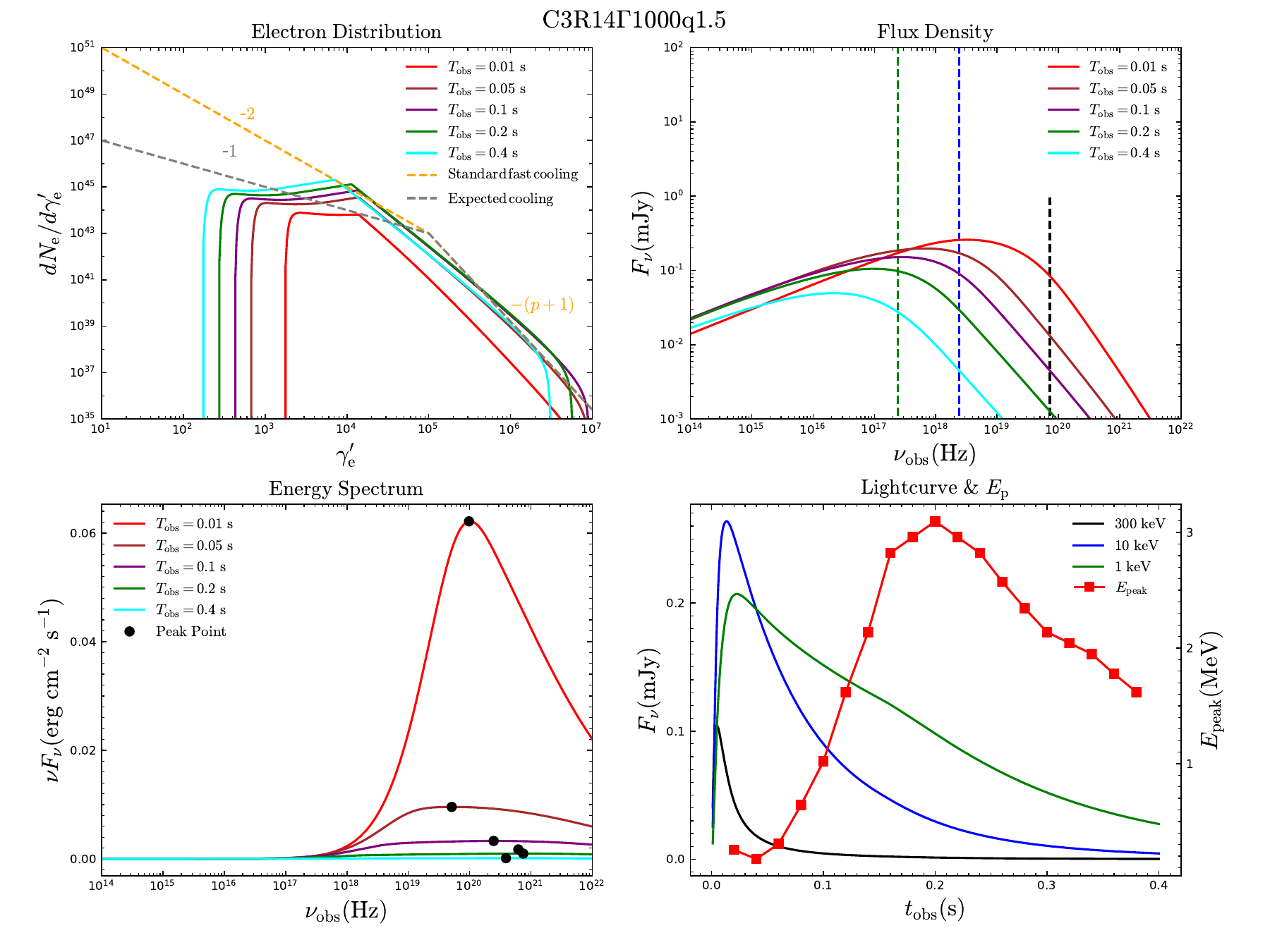}}\\
    \subfloat{\includegraphics[width=10.5cm,height=10.6cm]{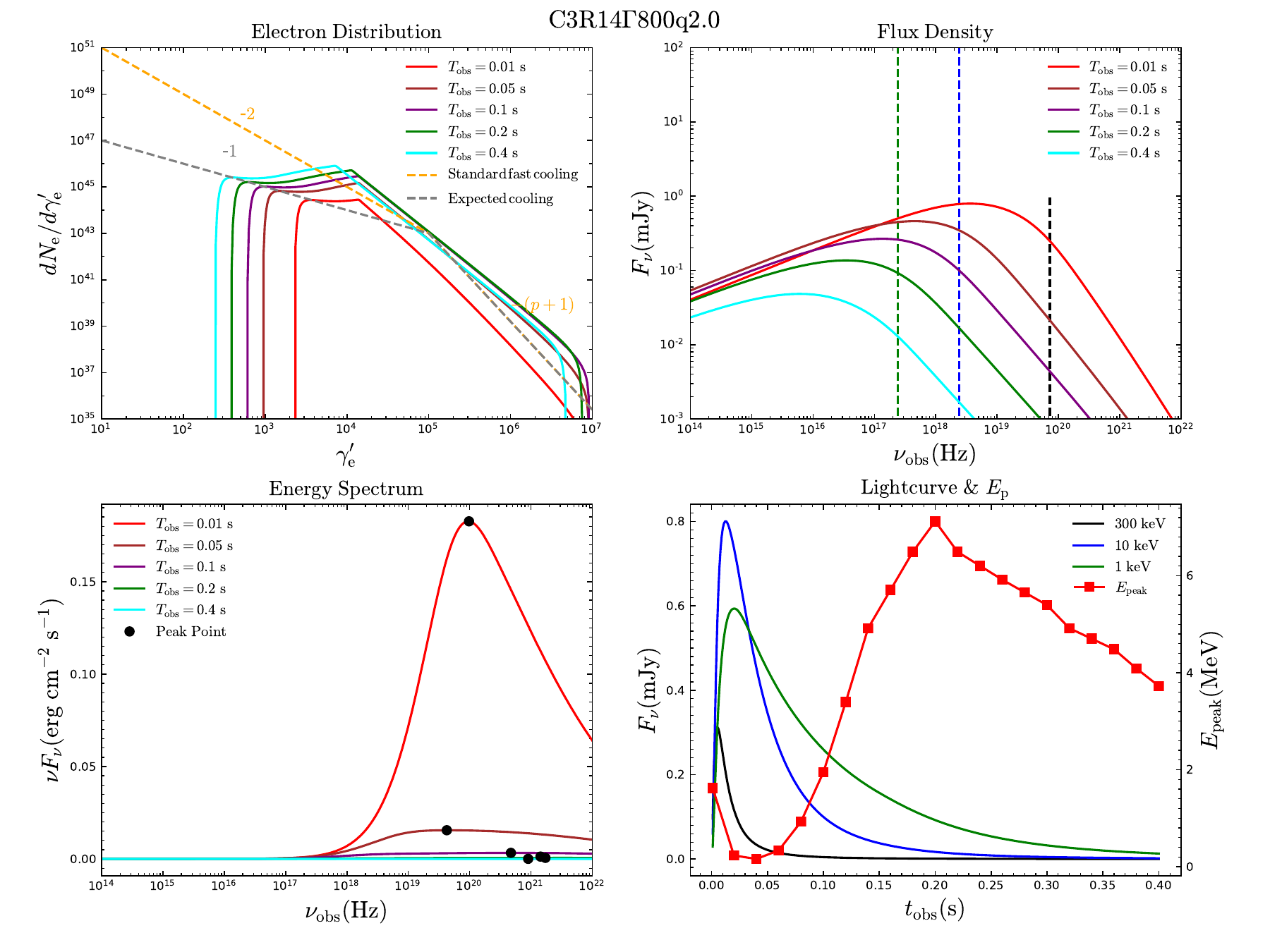}}
\end{adjustwidth}
\caption{ Numerical results for three cases shown in Table \ref{defq}.}
\label{differetq}
\end{figure*}

\appendix
\section{Formulations for synchrotron radiation model}
\label{app:flattau}
There are three main cooling mechanisms for relativistic electrons.
An electron will lose energy when it is moving in the magnetic field with a Lorentz factor $\gamma_{\mathrm{e}}^{\prime}$ \citep{Rybicki86},
\begin{equation}
\dot{\gamma}_{\text{e,syn}}^{\prime}=-\frac{\sigma_{T} B^{\prime 2} \gamma_{\mathrm{e}}^{\prime 2}}{6 \pi m_{\mathrm{e}} c},
\label{syn}
\end{equation}
where $\sigma_{T}$ is the Thomson cross-section.
Meanwhile, the electron undergoes an adiabatic cooling \citep{Uhm12,Geng14} in the moving jet,
\begin{equation}
\dot{\gamma}_{\mathrm{e}, \mathrm{adi}}^{\prime}=
\frac{1}{3} \gamma_{\mathrm{e}}^{\prime} \frac{d \ln n_{\mathrm{e}}^{\prime}}{d t^{\prime}}=-\frac{2}{3} \frac{\gamma_{\mathrm{e}}^{\prime}}{R} \frac{d R}{d t^{\prime}},
\label{adi}
\end{equation}
where $n_{\mathrm{e}}^{\prime}$ is the comoving electron number density.
Moreover, electrons would cool by scattering with self-emitted synchrotron photons, called the SSC process~\citep{Blumenthal70},
\begin{equation}
\begin{aligned}
\dot{\gamma}_{\mathrm{e}, \mathrm{SSC}}^{\prime}=&-\frac{1}{m_{\mathrm{e}} c^{2}} \frac{3 \sigma_{T} c}{4 \gamma_{\mathrm{e}}^{\prime 2}} \int_{\nu_{\min }^{\prime}}^{\nu_{\max }^{\prime}} \frac{n_{\nu^{\prime}} d \nu^{\prime}}{\nu^{\prime}} \\
& \times \int_{\nu_{\mathrm{ic}, \min }^{\prime}}^{\nu_{\mathrm{ic}, \max }^{\prime}} h \nu_{\mathrm{ic}}^{\prime} d \nu_{\mathrm{ic}}^{\prime} F(q, g),
\end{aligned}
\end{equation}
where $n_{\nu^{\prime}}$, $\nu^{\prime}$, and $\nu_{\mathrm{ic}}^{\prime}$ are the synchrotron seed photon spectrum, the frequency of the photon before scattering, and the frequency of the photon after scattering, respectively.
Here, $F(q, g)=2 q \ln q+(1+2 q)(1-q)+\frac{1}{2} \frac{(4 q g)^{2}}{1+4 q g}(1-q)$,
$g=\frac{\gamma_{\mathrm{e}}^{\prime} h \nu^{\prime}}{m_{\mathrm{e}} c^{2}}$,
$w=\frac{h \nu_{\mathrm{ic}}^{\prime}}{\gamma_{\mathrm{e}}^{\prime} m_{\mathrm{e}} c^{2}}$, and $q=\frac{w}{4 g(1-w)}$,
which means that the Klein-Nishina effect was fully taken into account.
The upper limit of the internal integral can be derived as $h \nu_{\mathrm{ic}, \max }^{\prime}=\gamma_{\mathrm{e}}^{\prime} m_{\mathrm{e}} c^{2} \frac{4 g}{4 g+1}$,
and the lower limit is $\nu_{\text {ic,min}}^{\prime}=\nu^{\prime}$.
To sum up the above, we could obtain the electron distribution by solving the continuity equation of electrons in energy space~\citep{Longair11},
\begin{equation}
\frac{\partial}{\partial t^{\prime}}\left(\frac{d N_{\mathrm{e}}}{d \gamma_{\mathrm{e}}^{\prime}}\right)+\frac{\partial}{\partial \gamma_{\mathrm{e}}^{\prime}}\left[\dot{\gamma}_{\mathrm{e}, \mathrm{tot}}^{\prime}\left(\frac{d N_{\mathrm{e}}}{d \gamma_{\mathrm{e}}^{\prime}}\right)\right]=Q\left(\gamma_{\mathrm{e}}^{\prime}, t^{\prime}\right),
\end{equation}
where $\dot{\gamma}_{\mathrm{e}, \mathrm{tot}}^{\prime}=\dot{\gamma}_{\mathrm{e}, \mathrm{syn}}^{\prime}+\dot{\gamma}_{\mathrm{e}, \mathrm{adi}}^{\prime}+\dot{\gamma}_{\mathrm{e}, \mathrm{SSC}}^{\prime}$.
More details on the numerical method could be found in \citet{Yabe01} and \citet{Geng18b}.

\section{Constraints of parameters}

Here, some constraints on parameters used in our scenario are derived.
The photosphere radius is estimated by~\citep{Meszaros00,Rees05,Deng14}
\begin{equation}
r_{\text{ph}} =(1-\beta)\frac{L\sigma_\text{T}}{4 \pi m_{\mathrm{p}} \beta c^{3} \Gamma},
\end{equation}
where $L$ is the luminosity of the shell, $m_{\mathrm{p}}$ is the mass of the proton, and $\beta$ is the dimensionless velocity of the jet.
For a range of the bulk Lorentz factor $\Gamma$ from 50 to 300,
the photosphere radius $r_\text{ph}$ varies from $10^{11}$~cm to $10^{13}$~cm,
where the radiation is described by the photosphere emission.
Meanwhile, we found that the scattering between electrons and photons is dramatic below $r_\text{ph}$
and our calculations are invalid under such large optical depth.
A more delicate model containing both the photosphere emission and the synchrotron radiation should be invoked in the future.
On other hand, when the radius $R_{0}$ is larger than $10^{16}$~cm,
synchrotron cooling always dominates the process and electrons are in the fast cooling regime.
Thus we focus on the initial radius ranging from $10^{14}$~cm to $10^{15}$~cm for the shell in this work.

The radiative cooling time for an electron of $\gamma_{\mathrm{e}}^{\prime}$ in the observer frame is~\citep{Sari98,Sari01}
\begin{equation}
t_{\mathrm{c}}=\frac{6 \pi m_{\mathrm{e}} c(1+z)}{\sigma_{T} B^{\prime 2} \gamma_{\mathrm{e}}^{\prime} \Gamma(1+Y)},
\label{tc}
\end{equation}
where $Y$ is the Compton parameter.
The fast-cooling condition of $t_{\mathrm{c}}\left(\gamma_{\mathrm{m}}^{\prime}\right) \leqslant t_{\rm obs}$ requires
\begin{equation}
B^{\prime} > 400~\Gamma_2^{1/2} \gamma_{\mathrm{m},3}^{\prime -1/2} R_{0,14}^{-1/2}\, \mathrm{G},
\end{equation}
where the convention $Q_x=Q/10^{x}$ in cgs units are adopted hereafter.
Since the extremely fast-cooling regime is not favored by observations,
we set the upper limit of $B^{\prime} < 400$ G in our calculations.
The range of is $\Gamma$ taken to be within [50,2000] according to
constraints from other observations in the literature~\citep{Lithwick01,Abdo09a}.
Furthermore, $\gamma_{\rm m}^{\prime}$ varies from $10^3$ to several $10^4$
in line with particle-in-cell simulations~\citep{Sironi09a,Sironi13,Guo14}.

\section{Effect of different $q$}
\label{app:flattau}

The value of the magnetic field decaying index ($q$) varies from 1 to 2 due to the geometry of
the shell and the structure of the magnetic field~\citep{Meszaros99,2001A&A...369..694S}.
Moreover, it was suggested that $q$ could be 0.6 in recent studies~\citep{Ronchini21}.
To investigate the effect of different $q$ on our results in the main text,
we adopt the other three possible values of $q$ (i.e., 0.5, 1.5, and 2) and perform numerical calculations for different cases as above.
We find that the results under the cooling process dominated by synchrotron cooling or SSC cooling
are weakly influenced by different $q$ since the absolute value of $B^{\prime}$ is more crucial than its decaying law.
For simplicity, only the results in the dominance of adiabatic cooling are exhibited here.
The model name follows the previous definition, i.e., ``$\text{CXRY}\Gamma \text{Z}$'' but with an additional parameter $q$,
and detailed parameters could be found in Table~\ref{defq}.

As shown in Fig. \ref{differetq}, in $\text{C3R14}\Gamma \text{1900}q0.5$, where $q=0.5$,
$E_{\rm p}$ shows a hard-to-soft evolution, owing to the slow decay of the magnetic field and the adiabatic cooling can not dominate at the early time.
In $\text{C3R14}\Gamma \text{1000}q1.5$, the evolution of the electron distribution obeys the adiabatic cooling case.
However, at an early stage, due to a large value of $q$, the synchrotron cooling rate decreases rapidly
the adiabatic cooling dominates soon and $E_{\rm p}$ rises.
At a late time, the observed lightcurve may even enter the decay phase before $R_{\rm off}$.
Also, the rapid decay of synchrotron intensity along with $R$ makes
the observed spectrum/$E_{\rm p}$ is dominated by the high latitude emission of the EATS.
Similarly, \citet{Ronchini21} find that the spectral evolution becomes dominated by the emission at larger angles for large values of $q$.
Therefore, a chromatic intensity-tracking evolution pattern of $E_{\rm p}$ is produced.
Moreover, a similar result is shown in $\text{C3R14}\Gamma \text{800}q2.0$.
These results support our conclusion on the crucial role of adiabatic cooling dominance for an intensity-tracking pattern.
On the other hand, a large $q$, accompanying with a dominated adiabatic cooling, would generate a chromatic intensity-tracking evolution of $E_{\rm p}$.
\begin{table*}
\caption{Parameters of numerical calculations for different $q$.}
\label{Table:results}
\centering
\begin{tabular}{c c c c c c c c}
\hline\hline
Model & $\Gamma$ & $\gamma^{\prime}_{m}$ & $B^{\prime}_{0}$    & $N^{\prime}_{\text {inj}}$  & $R_{0}$  & $q$   \\

      &          & $(10^{4})$    &($10^{2}$ G) &$(10^{47} \text {s}^{-1}$) & (cm)\\
\hline
$\text {C} 3 \text {R} 14  \Gamma 1900 $  & 1900  & 1.4   & 3     & 0.1  &  $10^{14}$   &$0.5$\\
$\text {C} 3 \text {R} 14  \Gamma 1000 $  & 1000  & 1.4   & 3     & 1.2  &  $10^{14}$   &$1.5$ \\
$\text {C} 3 \text {R} 14  \Gamma  800 $  &   800  & 1.4  & 3     & 5.8  &  $10^{14}$   &$2.0$\\
\hline
\end{tabular}
\label{defq}
\end{table*}

\section{Concerns on EATS}
\label{app:flattau}

By ignoring the EATS effect,
one could often estimate the observed flux density by using
\begin{equation}
F_{\nu_{\mathrm{obs}}}=\frac{(1+z) \Gamma P^{\prime}\left(\nu^{\prime}\left(\nu_{\mathrm{obs}}\right)\right)}{4 \pi D_{L}^{2}},
\label{eq:non-EATS}
\end{equation}
where $P^{\prime}$ is the synchrotron radiation power of all electrons at frequency $\nu^{\prime}=(1+z) \nu_{\mathrm{obs}} / D$,
$D=1 /[\Gamma(1-\beta \cos \theta)]$ is the Doppler factor,
and $\theta$ is the angle between the line of sight and the local radial direction.
When we take the EATS effect into account, the observed flux density now writes as~\citep{Geng16}
\begin{equation}
F_{\nu_{\mathrm{obs}}}=\frac{1+z}{4 \pi D_{L}^{2}} \int_{0}^{\theta_{j}}
P^{\prime}\left(\nu^{\prime}\left(\nu_{\mathrm{obs}}\right)\right) D^{3} \frac{\sin \theta}{2} d \theta,
\label{eq:EATS}
\end{equation}
where $\theta_{j}$ is the half-opening angle of the jet.

In general, the results from the integral of Equation (\ref{eq:EATS}) should degenerate to results from Equation (\ref{eq:non-EATS}) in practice.
However, this is not the case in scenarios invoking a starting radius $R_0$.
At the beginning of the radiation, the shell moves forward a little distance relative to its initial radius.
When we are performing the integral of Equation (\ref{eq:EATS}), only a small region contributes
to the observed flux since the high latitude locations of the full EATS are not filled with photons at all.
On the other hand, if we use Equation (\ref{eq:non-EATS}) and take $\theta$ to be 0,
it means that we have supposed all photons of the shell converge on the beam of light of sight.
This treatment would highly overestimate $F_{\nu_{\rm obs}}$ in comparison with that from Equation (\ref{eq:EATS}).
The approximate formulation for $F_{\nu_{\rm obs}}$ in Equation (\ref{eq:non-EATS}) is therefore invalid
in this work, which should also be noted in future relevant works.

\end{document}